\documentclass[%
 reprint,
superscriptaddress,
 amsmath,amssymb,
 aps,
]{revtex4-2}
\usepackage{graphicx}
\usepackage{dcolumn}
\usepackage{bm}
\usepackage{ulem}
\usepackage{braket}
\usepackage[english]{babel}
\usepackage{physics}
\usepackage[dvipsnames]{xcolor}
\usepackage{hhline}
\usepackage{cancel}

\usepackage{amsmath}

\usepackage{amssymb}
\usepackage{graphicx}
\usepackage{braket}
\usepackage{babel}
\usepackage{hhline}
\makeatletter
\usepackage[dvipsnames]{xcolor}

\usepackage{mathtools}

\makeatletter

\@ifundefined{textcolor}{}
{
 \definecolor{BLACK}{gray}{0}
 \definecolor{WHITE}{gray}{1}
 \definecolor{RED}{rgb}{1,0,0}
 \definecolor{GREEN}{rgb}{0,1,0}
 \definecolor{BLUE}{rgb}{0,0,1}
 \definecolor{CYAN}{cmyk}{1,0,0,0}
 \definecolor{MAGENTA}{cmyk}{0,1,0,0}
 \definecolor{YELLOW}{cmyk}{0,0,1,0}
 }

\usepackage{bm}\@ifundefined{definecolor}{\usepackage{color}}{}

\usepackage{textcomp}
\usepackage{color}

\usepackage{babel}
\newcommand\matthree[9]{%
     \begin{pmatrix*}[r]
         #1 & #2 & #3 \\ 
         #4 & #5 & #6 \\ 
         #7 & #8 & #9
     \end{pmatrix*}}
     
\begin{document}

\title{Quantum Information Scrambling on a Superconducting Qutrit Processor}

\author{M.S. Blok}
\thanks{These authors contributed equally to this work.}
\affiliation{Lawrence Berkeley National Laboratory, Berkeley, CA 94720, United States of America}
\affiliation{Department of Physics, University of California, Berkeley, California 94720 USA}
\author{V.V. Ramasesh}
\thanks{These authors contributed equally to this work.}
\affiliation{Lawrence Berkeley National Laboratory, Berkeley, CA 94720, United States of America}
\affiliation{Department of Physics, University of California, Berkeley, California 94720 USA}
\author{T. Schuster}
\affiliation{Department of Physics, University of California, Berkeley, California 94720 USA}
\author{K. O'Brien}
\affiliation{Department of Physics, University of California, Berkeley, California 94720 USA}
\author{J.M. Kreikebaum}
\affiliation{Lawrence Berkeley National Laboratory, Berkeley, CA 94720, United States of America}
\affiliation{Department of Physics, University of California, Berkeley, California 94720 USA}
\author{D. Dahlen}
\affiliation{Lawrence Berkeley National Laboratory, Berkeley, CA 94720, United States of America}
\affiliation{Department of Physics, University of California, Berkeley, California 94720 USA}
\author{A. Morvan}
\affiliation{Lawrence Berkeley National Laboratory, Berkeley, CA 94720, United States of America}
\affiliation{Department of Physics, University of California, Berkeley, California 94720 USA}
\author{Beni Yoshida}
\affiliation{Perimeter Institute for Theoretical Physics, Waterloo, Ontario N2L 2Y5, Canada}
\author{N.Y. Yao}
\affiliation{Department of Physics, University of California, Berkeley, California 94720 USA}
\author{I. Siddiqi}
\affiliation{Lawrence Berkeley National Laboratory, Berkeley, CA 94720, United States of America}
\affiliation{Department of Physics, University of California, Berkeley, California 94720 USA}

\date{\today}

\begin{abstract}
  The dynamics of quantum information in strongly-interacting systems, known as quantum information \textit{scrambling}, has recently become a common thread in our understanding of black holes, transport in exotic non-Fermi liquids, and many-body analogs of quantum chaos.
  To date, verified experimental implementations of scrambling have focused on systems composed of two-level qubits.
  Higher-dimensional quantum systems, however, may exhibit different scrambling modalities  and are predicted to  saturate conjectured speed limits on the rate of quantum information scrambling.
  We take the first steps toward accessing such phenomena, by realizing a quantum processor based on superconducting qutrits (three-level quantum systems).
  We demonstrate the implementation of universal two-qutrit scrambling operations and embed them in a five-qutrit quantum teleportation protocol.
  Measured teleportation fidelities, $F_{\textrm{avg}} = 0.568 \pm 0.001$, confirm the presence of scrambling even in the presence of experimental imperfections and decoherence.
  Our teleportation protocol, which connects to recent proposals for studying traversable wormholes in the laboratory, demonstrates how quantum technology that encodes information in higher-dimensional systems can exploit a larger and more connected state space to achieve the resource efficient encoding of complex quantum circuits.
\end{abstract}

\maketitle

\section{\label{sec:introduction} INTRODUCTION}

\begin{figure*}
\includegraphics[scale=1]{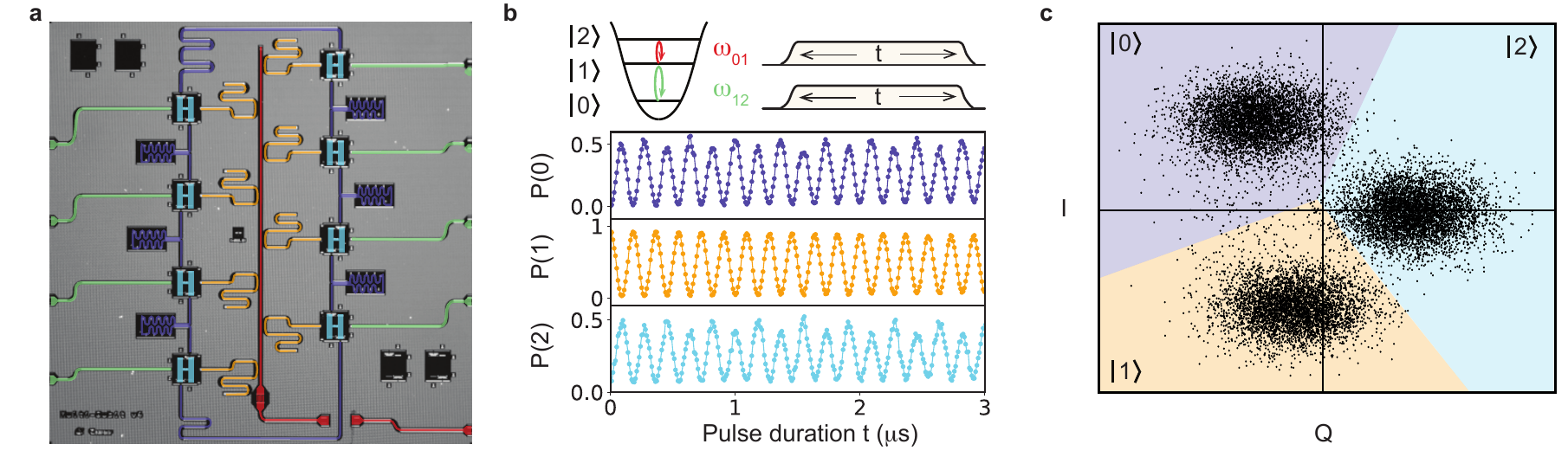}\caption{ \textbf{Superconducting qutrit processor | a,} Optical micrograph of the five-transmon processor used in this experiment.  Transmon circuits (light blue) couple to an integrated Purcell-filter and readout bus (red) via individual linear resonators (gold), enabling multiplexed state measurement.  Exchange coupling between nearest-neighbor transmons is mediated by resonators (purple), while microwave drive lines (green) enable coherent driving of individual qubits. \textbf{b,} Coherent Rabi dynamics of a single qutrit induced by simultaneous microwave driving at frequencies $\omega_{01}$ and $\omega_{12}$.  Achievable Rabi frequencies are in the range of tens of MHz, three orders of magnitude faster than decoherence timescales.  \textbf{c,} Example single-shot readout records of an individual qutrit, generally achievable with fidelities above 0.95.  This is largely limited by decay during readout. 
\label{fig:fig2}}
\end{figure*}

While the majority of current generation quantum processors are based on qubits, qutrit-based (and more generally qudit-based \cite{luo2014universal}) systems have long been known to exhibit significant advantages in the context of quantum technology: they have been touted for their small code-sizes in the context of quantum error correction~\cite{Campbell_2014}, high-fidelity magic state distillation~\cite{campbell2012magic}, and more robust quantum cryptography protocols~\cite{bruss2002optimal, bechmann2000quantum}. To date, advantages of individual qutrits have been explored experimentally in fundamental tests of quantum mechanics~\cite{lapkiewicz_experimental_2011} and in certain quantum information protocols ~\cite{fedorov_implementation_2012,kurpiers_deterministic_2018,rosenblum_fault-tolerant_2018,dai_demonstration_2018}, while entanglement between qutrits has been demonstrated in probabilistic photonic systems~\cite{Luo_2019,imany_npjQI_2019}.
A qutrit platform capable of implementing deterministic high-fidelity gates would be a powerful tool for both quantum simulation and information processing.  In this work, we develop a prototypical multi-qutrit processor based on superconducting transmon circuits (Fig.~\ref{fig:fig2}), and---as a proof-of-principle demonstration---use it to (i) perform a maximally-scrambling two-qutrit unitary and (ii) verify its scrambling behavior using a five-qutrit quantum teleportation algorithm.  

Quantum scrambling, the subject of much recent interest, is the quantum analogue of chaotic dynamics in classical systems:  scrambling describes many-body dynamics which, though ultimately unitary, scatter initially-localized quantum information across all of the system's available degrees of freedom~\cite{Shenker_2014, Hayden_2007, Sekino_2008}.  Just as chaotic dynamics enable thermalization in closed classical systems, scrambling dynamics enable thermalization of isolated many-body quantum systems by creating highly-entangled states.  In general, it is difficult to determine whether arbitrary many-body Hamiltonians lead to scrambling dynamics; this is one area where future quantum processors could offer experimental guidance.  Our realization of verifiable scrambling behavior in a multi-qutrit system is a step towards this goal.  

In particular, a quantum processor can, in principle, directly measure the scrambling-induced spread of initially localized information via the decay of so-called out-of-time-ordered correlation functions (OTOCs)~\cite{larkin1969quasiclassical, Shenker_2014, Roberts_2015, Maldacena_2016, Hosur_2016,Garttner_2017, PhysRevX.7.031011, PhysRevLett.120.070501, Lepoutre_2019}.  This capability was recently demonstrated in a qubit-based system, using a seven-qubit teleportation algorithm analogous to the five-qutrit protocol we implement in this work~\cite{yoshida2017efficient,Yoshida_2019,Landsman_2019}.  A key property of the teleportation-based protocol is that it enables verification of scrambling behavior even in the face of decoherence and experimental imperfection. 

The superconducting qutrit processor we develop here features long coherence times; multiplexed readout of individual qutrits; fast, high-fidelity single-qutrit operations; and two types of two-qutrit gates for generating entanglement.  Using this gateset on our processor, we construct a maximally scrambling qutrit unitary and characterize it using quantum process tomography.  Finally, to demonstrate the ability of our platform to perform genuine multi-qutrit algorithms, we perform a five-qutrit teleportation protocol which serves as an additional verification of genuine two-qutrit scrambling behavior.  This protocol is inspired by the Hayden-Preskill variant of the black-hole information paradox~\cite{Hayden_2007,yoshida2017efficient}. While we have chosen quantum scrambling as a demonstration of our processor, our work opens the door more broadly to the experimental study of quantum information processing utilizing qutrit-based logic.  

\section*{II. QUTRIT PROCESSOR}

Nearly all current quantum processors are based on collections of the simplest possible quantum system: qubits, consisting of two states per site. In contrast, qudits -- featuring $d > 2$ states -- can store an exponentially greater amount of information compared to qubits and can therefore, in certain cases, implement quantum algorithms using smaller systems and fewer multi-site entangling gates       ~\cite{gokhale_arxiv_2019, bullock_asymptotically_2005}. In cavity QED experiments, continuous variable quantum information has been encoded~\cite{sayrin_real-time_2011,naik_NatComm_2017}, protected~\cite{ofek_extending_2016} and entangled ~\cite{gao_Nature_2019} in linear cavity modes, by coupling them to an artificial non-linear atom. Encoding discrete quantum information in an intrinsically non-linear  system has the advantage that energy transitions have distinct frequencies that can be driven individually and with faster gates. 

For qutrit systems specifically, many quantum information protocols have been proposed that would yield a significant advantage over qubits~\cite{Campbell_2014,campbell2012magic,bruss2002optimal, bechmann2000quantum,baekkegaard_realization_2019}. Qutrits have been successfully realized in various physical degrees of freedom including the polarization state of multiple photons \cite{lanyon_manipulating_2008}, spin $S=1$ states of solid-state defect centers \cite{dolde_high-fidelity_2014}, hyperfine states in trapped atoms and ions \cite{sadler2006spontaneous,senko_realization_2015} and the lowest energy states of an anharmonic oscillator in superconducting circuits \cite{bianchetti_PRL_2010}. However, no platform to date has demonstrated a deterministic, universal two-qutrit gate.

Our superconducting qutrit processor features eight transmons~\cite{Koch_2007, Schreier_2008} connected in a nearest-neighbor ring geometry (Fig.~\ref{fig:fig2}a) of which we use five (denoted $Q_1, \ldots, Q_5$) to realize quantum circuits. Each transmon encodes a single qutrit and is coupled both to a dedicated microwave control line (for performing gates, Fig.~\ref{fig:fig2}b) and its own readout resonator (for state measurement, Fig.~\ref{fig:fig2}c). Transmons are quantum nonlinear oscillators that can be operated as qubits, using only their two lowest-lying energy states $\ket{0}$ and $\ket{1}$. While their higher energy states ($\ket{2}$, $\ket{3}$, etc.) in principle enable transmons to be operated as higher-dimensional qutrits or qudits \cite{bianchetti_PRL_2010} , the experimental implementation of multi-qutrit algorithms is challenging due to a lack of two-qutrit entangling gates and increased noise associated with the higher transmon states. 

\subsection*{A. High-fidelity single qutrit operations}

In order to implement high-fidelity single-qutrit operations in transmons, one must overcome multiple sources of noise and coherent errors that are naturally introduced upon including the $\ket{2}$ state of transmon in the computational subspace.  The primary such sources include: 

($i$) \textit{Relaxation due to spontaneous emission}---Energy relaxation or $T_1$ processes in transmon qubits cause unwanted, incoherent transitions from the $\ket{1}$ state to the $\ket{0}$ state, and arise from lossy interfaces or electromagnetic channels.  For transmon qutrits, the addition of the $\ket{2}$ state introduces another decay channel, from $\ket{2}\rightarrow\ket{1}$~\footnote{In principle, one would also have to consider the transition from $\ket{2}\rightarrow\ket{0}$, but this channel is largely suppressed, due to parity, in transmons.}. Due to bosonic enhancement, in which the spontaneous emission rate scales linearly with photon number, the time constant associated with $\ket{2}\rightarrow\ket{1}$ decay is roughly half that of the $\ket{1}\rightarrow\ket{0}$ transition. 

To address this, we use a high-quality fabrication process and careful microwave engineering to minimize the effects of  energy relaxation.  The fabrication recipe, detailed in Appendix \ref{app:Fab}, is optimized to remove lossy oxide layers at the substrate-metal and metal-metal interfaces through three separate buffered-oxide etches and an ion-mill step.  In addition, finite-element simulations are used to identify and mitigate each loss channel associated with microwave radiation.  Specific mitigation strategies include: an integrated, broadband Purcell filter suppressing leakage into the readout bus (Fig.~\ref{fig:fig2}a); readout and bus resonator geometries designed to shift the frequencies of lossy higher modes away from those of qutrits and resonators of band; and extensive wire-bonding connecting the ground planes on either sides of the readout bus, coupling resonators, readout resonators, and control lines.  As a results of these techniques, the average $T_1$ times on the chip are $56.0 \pm 10~\mu$s for the $\ket{1}\rightarrow\ket{0}$ transition, and $34.8 \pm 4~\mu$s for the $\ket{2}\rightarrow\ket{1}$ transition.  

($ii$) \textit{Dephasing due to the charge-sensitivity of higher transmon levels}---The transmon was originally developed to reduce the dependence of energy levels on offset charge noise.  While typical values of transmon parameters (i.e. the Josephson energy $E_J$ and the charging energy $E_C$)  result in low charge-noise sensitivity in the qubit subspace, the charge dispersion increases rapidly with increasing energy levels~\cite{PhysRevLett.114.010501, Koch_2007}.  For qutrits specifically, the charge dispersion of the $\ket{2}$ state is at least an order of magnitude greater than that of the $\ket{1}$ state, resulting in a charge-limited dephasing time ten times lower than that of the qubit subspace. 

To mitigate charge noise in the $\ket{2}$ state, we tune the transmon parameters even further than typical into the ``transmon regime'': specifically, we choose a ratio of Josephson and charging energies $E_J/E_C \approx 73$ (typically, this ratio is near 50 for transmon qubits~\cite{Barends_2014, Kandala_2019,Salath__2015}).  This reduces the charge dispersion of the $\ket{2}$ and $\ket{1}$ states to 12 kHz and 250 Hz, respectively.  As a result, we realize dephasing times, averaged over the five qutrits, of $T_\textrm{2}^{*}= (39 \pm 21) ~\mu$s [$(14 \pm 5)~\mu$s] for the $\ket{0} \rightarrow \ket{1}$ [$\ket{1} \rightarrow \ket{2}$] transitions, which can be further extended with a Hahn-echo to $T_\textrm{2,echo}=(61.2 \pm 11)~\mu$s [$(28 \pm 5)~\mu$s].  

($iii$) \textit{Increased cross-talk due to frequency crowding}--- If pulses, intended to apply unitary operations on a single qutrit, are not well localized in space to the desired qutrit, they can induce unintended unitary operations on nearby qutrits.  This `cross-talk' is increasingly troublesome when including the $\ket{2}$ state, as the frequency spectrum of state transitions becomes more crowded due to the inclusion of the $\ket{1}\rightarrow\ket{2}$ transition frequencies. 

On our processor, we find significant cross-talk between the microwave drive lines for most transmons.  When driving Rabi oscillations on a given qutrit, this cross-talk has two unwanted effects:
\begin{enumerate}
    \item All other qutrits are off-resonantly driven.  Depending on the relative frequencies between the qutrits, this can manifest either as an unwanted shift in state populations or as an AC Stark shift.  
    \item Microwave fields leaking onto neighboring qutrit(s) will result in an unwanted cross-resonance interaction, making the desired Rabi frequency dependent on the state of the neighboring qutrit(s)~\cite{RigettiDevoret}.
\end{enumerate}

For a given drive frequency, our cross-talk can be characterized in terms of a five-by-five complex-valued matrix $C(\omega)$, which relates the field amplitudes $\vec{o}(\omega)$ seen by each of the five qutrits to the input field amplitudes $\vec{i}(\omega)$ on each drive line: $\vec{o}(\omega)=C(\omega) \vec{i}(\omega)$. While this cross-talk matrix strongly depends on frequency, it does not exhibit any nonlinearity at the powers used in the experiment. 
Therefore, we compensate for the cross-talk by inverting the matrix $C(\omega)$ at each frequency to yield linear combinations of drive lines which result in non-zero fields only at a single qutrit.

To measure the cross-talk matrix $C(\omega)$ of our system, we drive two control lines simultaneously, and, for each pair of driven control lines, determine the relative amplitudes and phases of driving that cancel the resulting field at each qutrit on the chip.
Depending on the relative frequencies between the drive field and the qutrit transition, we use either an AC Stark shift or a Rabi oscillation as the diagnostic of an unwanted microwave field. This measurement is repeated for each of ten drive frequencies of interest (i.e. the $|0\rangle \leftrightarrow|1\rangle $ and $|1\rangle \leftrightarrow|2\rangle$ transition frequencies of all five qutrits), each pair of lines, and each qutrit.  

While this method proved effective for our system, it can become prohibitively measurement-intensive for future processors with hundreds of qudits. Looking forward, in order to scale to such systems it will be crucial to pinpoint and mitigate the source of this cross-talk at the hardware level.  

With the above three issues---relaxation, dephasing, and cross-talk---taken care of, we now turn to our realization of high-fidelity single-qutrit gates.  In qutrits, these require the combination of a microwave pulse (to perform the rotation in the relevant subspace) and a virtual phase-gate.  This phase gate is necessary because a microwave field that addresses, e.g. the $\ket{0}\rightarrow\ket{1}$ transition, also generates a relative phase $e^{i (\phi_G+\phi_D)} \ket{2}$ on the idling state, where $\phi_G$ is the geometric (Berry) phase associated with the rotation in the $\{\ket{0}, \ket{1}\}$ subspace and $\phi_D$ is the dynamical phase due to the AC Stark shift of the idling $\ket{2}$ state.  These phases represent logical Z errors in the computational subspace, which we measure and correct.

Single-qutrit gates in our system are performed within 30~ns (see Apendix \ref{app:QutritGates} for the universal set of gates used in our experiment). 
We characterize them by performing  qubit-based randomized benchmarking in two different qubit subspaces~\cite{proctor2017randomized}.
These yield fidelities  on par with state-of-the-art qubit processors: $f_{01} =0.9997 \pm 0.0001$ and $f_{12} = 0.9994 \pm 0.0001$ for gates within the $\{ \ket{0}, \ket{1} \}$ and $\{ \ket{1}, \ket{2} \}$ subspace, respectively. While this benchmarking method demonstrates that single-qutrit coherence is no longer a major bottleneck for high-fidelity operations, it should be noted that this method is not sensitive to certain sources of errors, including phase errors in the idle state and multi-qutrit errors. In the future, a full characterization of qutrit operations will require the development of genuine qutrit randomized benchmarking protocols.

Multiplexed, individual qutrit readout is performed dispersively via a linear readout resonator coupled to each transmon.  As shown in Fig.~\ref{fig:fig2}c, with a suitable choice of the readout frequency and amplitude all three states of the transmon can be resolved in a single shot.  Averaged over all qutrits, our readout fidelity is $F_{\textrm{avg}} = 0.96 \pm .02$.  In ensemble measurements (e.g. when performing tomography), we can correct for this readout imperfection by storing a `confusion matrix' $M_{ij}$ of the conditional probabilities to measure a state $\ket{i}$ given an actual state $\ket{j}$. Applying its inverse to the measurement results allows one to infer the actual state populations. 

\subsection*{B. Two-qutrit entanglement}

Having outlined our implementation of high-fidelity single-qutrit operations, we next demonstrate two methods for generating controllable two-qutrit entangling gates in our system: the first based on the cross-resonance interaction~\cite{Chow_2011, RigettiDevoret}, and the second based on the longitudinal, or cross-Kerr, interaction.  Both of these operate on neighboring qutrits, giving our full eight-qutrit processor an effective ring geometry; restricting to five qutrits, this reduces to a linear geometry.  The two types of interaction give rise to two distinct forms of entangling gates: the cross-resonance interaction leads (up to a phase) to a conditional subspace rotation, while the cross-Kerr interaction leads to a controlled-SUM gate---the qudit analogue of the controlled-NOT gate. We now describe both of these gates in detail.  

\begin{figure}[ht]
\includegraphics[scale=1]{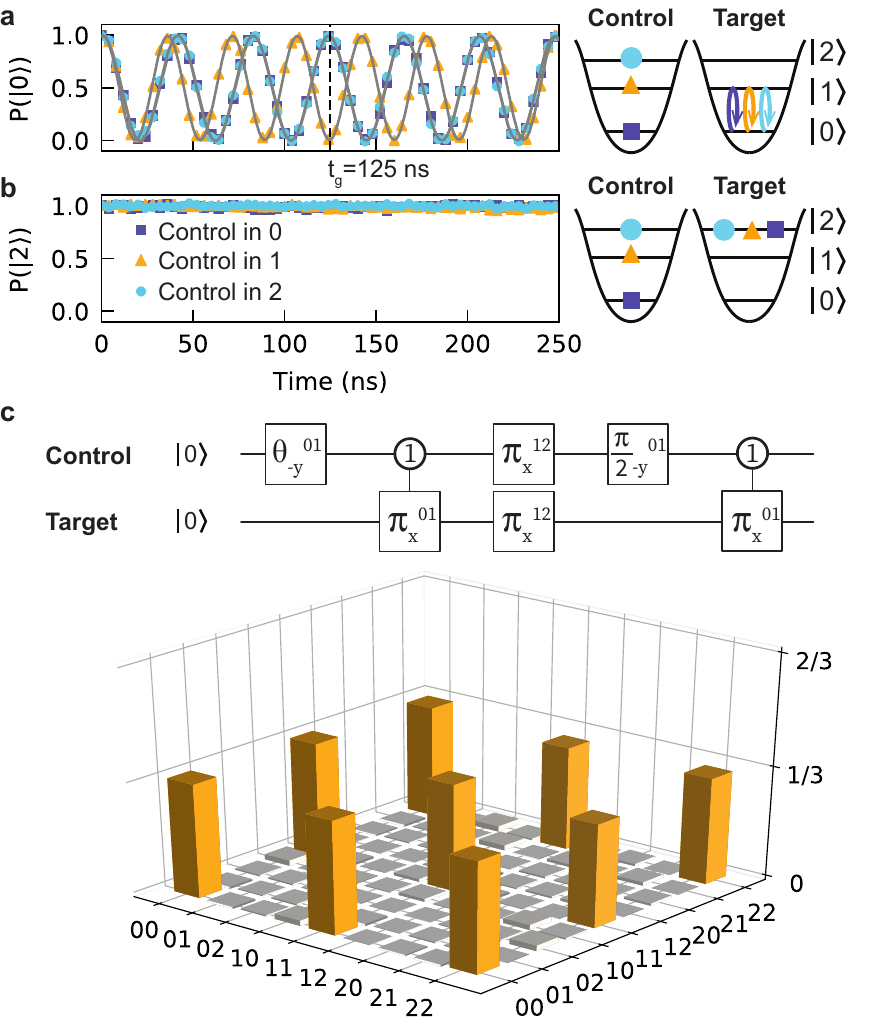}\caption{ \textbf{Two-qutrit EPR pair generation via the cross-resonance interaction | a} Nearest-neighbor qutrits coupled by an exchange interaction can be entangled via the cross-resonance effect, where one qutrit (the control) is microwave-driven at the 
$\ket{0}\leftrightarrow\ket{1}$ transition  frequency of the other (the target).  Resulting Rabi oscillations of the target qutrit exhibit a Rabi frequency dependent on the state of the control qutrit.  Here, we drive the control with a field whose amplitude is chosen to make the Rabi frequencies corresponding to control states $\ket{0}$ and $\ket{2}$ identical, resulting in a unitary operation which, after 125 ns, interchanges states $\ket{0}$ and $\ket{1}$ of the target qutrit when the control qutrit is the $\ket{1}$ state.  \textbf{b,} When the target qutrit is in the $\ket{2}$ state, the cross-resonance interaction is off-resonant and does not affect the population.  \textbf{c (top),} Sequence used to prepare an EPR pair with two applications of the cross-resonance gate. $\pi_x^{01}$ represents be a rotation of $\pi$ about the x axis in the $\{\ket{0},\ket{1}\}$ subspace of the qutrit. The number in the circle for the two-qutrit gate represents the condition (state of the control qutrit) for which a gate is applied on the target. (bottom) The density matrix (reconstructed via state tomography) of the resulting EPR pair, with a state fidelity of $F_\textrm{EPR}= 0.98 \pm 0.002$.
\label{fig:CRGate}}
\end{figure}

\textit{Cross-resonance gate}---The cross-resonance interaction has been used extensively in superconducting qubit systems to generate high-fidelity controlled-NOT gates~\cite{Sheldon_2016, Kandala_2017, jurcevic2020demonstration}. 
The origin of the interaction is a weak transverse coupling between two qubits that are highly off-resonant such that the coupling is almost negligible. When one of the two systems (the control) is driven at the frequency of the other (the target), the combination of the coupling and the drive induces Rabi oscillations on the target qubit, with a Rabi frequency that depends on the state of the control qubit.  
This results in two Rabi frequencies, $\omega_0$ and $\omega_1$, for the two control qubit states, $\ket{0}$ and $\ket{1}$, both of which also depend on the drive power. This interaction is naturally described by the Hamiltonian 
\begin{equation}
H^{\textrm{qubit}}_{cr} / \hbar= \omega_{0} \dyad{0} \otimes \sigma_x + \omega_1 \dyad{1}\otimes \sigma_x.
\end{equation}
Higher-order deviations from this Hamiltonian can largely be mitigated with dynamical decoupling, local rotations, or pulse shaping.  
This Hamiltonian is locally equivalent to a $\sigma_z \otimes \sigma_x$ interaction, and periodically entangles and disentangles qubits at a rate $\Delta \omega = |\omega_0 - \omega_1|$.  When the two oscillations are completely out of phase ($t\Delta \omega = \pi$), the resulting unitary
\begin{equation}
U_{\textrm{CNOT}}= \dyad{0}\otimes I + \dyad{1}\otimes \sigma_x
\end{equation}
is locally equivalent to a controlled-NOT gate.

On our chip, the cross-resonance interaction arises from the static capacitive coupling between neighboring transmons.
To extend the cross-resonance gate to \textit{qutrit} systems, we must consider how this interaction behaves in the full two-qutrit Hilbert space.  
We will show that we can realize a similar gate to that of qubit systems, where we swap the $\ket{0}$ and $\ket{1}$ populations of the target qutrit conditional on the control qutrit being in the $\ket{1}$ state.  

The dynamics of the two-qutrit cross-resonance interaction are shown in Fig.~\ref{fig:CRGate}(a-b). Compared to qubits, we have two additional considerations: the effect of the cross-resonance drive on the $\ket{2}$ state of the target qutrit, and the Rabi oscillations induced on the target qutrit by the $\ket{2}$ state of the control qutrit.  The first effect results in an overall phase being applied to the $\ket{2}$ state of the target via the AC Stark shift (Fig.~\ref{fig:CRGate}b), since the effective Rabi drive is off-resonant with the $\ket{1}\rightarrow\ket{2}$ transitions of the target qutrit (in contrast to the $\ket{0}\rightarrow\ket{1}$ transition, with which it is resonant). The second effect extends the qubit cross-resonance Hamiltonian to qutrits according to
\begin{eqnarray}
\nonumber
H^{\textrm{qutrit}}_{cr} / \hbar&=& \omega_{0} \dyad{0} \otimes \pi_x^{01} + \omega_1 \dyad{1}\otimes \pi_x^{01}  \\
&& + \, \omega_2 \dyad{2}\otimes \pi_x^{01},\label{eqn:CRHamiltonian}
\end{eqnarray}
where we have added the (drive-dependent) frequency $\omega_2$, which induces Rabi oscillations in the target qutrit when the control qutrit is in the $\ket{2}$ state. 

In the above Hamiltonian, $\pi_x^{01}$ is the $3 \times 3$ Gell-Mann matrix that couples the $\ket{0}$ and $\ket{1}$ states, and does nothing to the $\ket{2}$ state. We emphasize that this model for the interactions, while physically motivated and consistent with our experimental observations, is only approximate. Much like for the two-qubit cross-resonance gate~\cite{Tripathi_2019,Magesan_2020}, further study is needed to develop a precise model of the dynamics.

The qutrit cross-resonance interaction Eq.~(\ref{eqn:CRHamiltonian}) involves three distinct Rabi frequencies, corresponding to the three states of the control qutrit (Fig.~\ref{fig:CRGate}a).  At low drive power, these frequencies scale linearly with the drive amplitude, while at higher powers the dependence becomes nonlinear.  We utilize this nonlinearity to choose a drive power such that two of the Rabi frequencies are equal, $\omega_0 = \omega_2$.  Next, similar to the qubit implementation, we choose the gate time $t_g$ according to $t_g |\omega_0 - \omega_1| = \pi$. Up to state-dependent phases, this gives rise to a conditional-$\pi$ gate (also known as a generalized controlled-X gate~\cite{di2011elementary}):
\begin{eqnarray}\nonumber
U_{\textrm{C}\pi}&=&(\dyad{0} + \dyad{2})\otimes I \\ 
&& + \dyad{1} \otimes (\dyad{0}{1} + \dyad{1}{0} + \dyad{2}).
\end{eqnarray}
Our experiment features a frequency difference $\omega_0 - \omega_1 = 4$ MHz and a corresponding gate time $t_g = 125$ ns.  During this gate, simultaneously with the cross-resonance drive applied to the control qutrit, we apply a concurrent drive to the target qutrit which ensures that it ends up in a computational basis state at $t_g$.  This target drive, with Rabi frequencies in the tens of MHz, has the additional benefit of suppressing dephasing in the target qutrit's $\{\ket{0}, \ket{1}\}$ subspace during the gate~\cite{fanchini_continuously_2007}. 

In order to benchmark the performance of the conditional-$\pi$ gate, we use two applications of the gate to create a maximally entangled two-qutrit EPR pair:
\begin{equation}
\ket{\textrm{EPR}} = \frac{\ket{00} + \ket{11} + \ket{22}}{\sqrt{3}}.
\end{equation}
We measure the success of EPR preparation using qutrit state tomography~\cite{bianchetti_PRL_2010}, enabled by our high-fidelity local qutrit pulses and qutrit readout. From this we reconstruct the density matrix of our final state (Fig.~\ref{fig:CRGate}c) and compute an EPR fidelity of $F_\textrm{EPR}= 0.98 \pm 0.002$, which we observe to be mostly limited by decoherence. 

Finally, while the conditional-$\pi$ gate is fast and capable of generating high-fidelity entangled states, it acts only in a two-qubit subspace of the full two-qutrit Hilbert space.
Consequently, despite the conditional-$\pi$ gate being universal for qutrit computation when combined with arbitrary single qutrit gates~\cite{di2011elementary}, implementing general two-qutrit unitaries will require a high number of such gates.

This can be undesirable given the finite error associated with each gate.
To address this, in what follows we introduce a \emph{second} qutrit entangling gate, the controlled-SUM gate, which we will implement via the cross-Kerr interaction instead of the cross-resonance interaction.
Ultimately, we will use both gates to perform the full five-qutrit teleportation protocol, applying the conditional-$\pi$ gate to EPR preparation/measurement and the more powerful controlled-SUM gate to a complex two-qutrit scrambling operation.

\begin{table}
\centering
 \begin{tabular}{c c c c c} 
\hline  & ~$Q_1$/$Q_2$~ & ~$Q_2$/$Q_3$~ & ~$Q_3$/$Q_4$~  & ~$Q_4$/$Q_5$~ \\
\hhline{=====}
$\alpha_{11}$ & {-279} & {-138} & {-276} & {-262} \\
\hline $\alpha_{12}$ & {160} & {158} & {-631} & {-495} \\
\hline $\alpha_{21}$ & {-528} & {-335} & {243} & {-528} \\
\hline $\alpha_{22}$ & {-743} & {-342} & {-748} & {-708} \\
\hline
\end{tabular}
\caption{Measured cross-Kerr interaction strengths between nearest-neighbor pairs of transmons, in units of kHz.}
\label{Table:interactions}

\end{table}

\emph{Cross-Kerr gate}---As neighboring qutrits in our processor are capacitively coupled to one another, they are subject to an ``always-on'' cross-Kerr entangling interaction, which we harness to perform controlled-phase and controlled-SUM gates (the two are locally equivalent). The always-on cross-Kerr interaction, analogous to the dispersive interaction between a qubit and cavity, is modeled by the dispersive cross-Kerr Hamiltonian 
\begin{eqnarray}
    H_{\mathrm{cK}} / \hbar=\alpha_{11} \dyad{11} + \alpha_{12} \dyad{12} + \\
    \nonumber
    \alpha_{21}\dyad{21} + \alpha_{22}\dyad{22},
\end{eqnarray}
where we work in a rotating frame that cancels the Hamiltonian's action on the $\ket{0j},\ket{i0}$ states.
The cross-Kerr coefficients $\alpha_{ij}$ represent the rate at which phases accumulate on the $\ket{ij}$ state during idling.
Their values are sensitive to both the magnitude of the capacitive coupling between transmons as well as the transmon spectra.
While in principle one could design a processor with pre-specified values of $\alpha_{ij}$, this is impractical given current technology because the transmon spectra depend on the critical current of their Josephson junction, which cannot be precisely set in fabrication~\cite{Kreikebaum2019}.  
To maintain robustness against such variability, we therefore design the cross-Kerr gate to work regardless of the particular values of the coefficients. 
For reference, these values for our particular processor are shown in Table~\ref{Table:interactions}.

We will now demonstrate how to combine a generic cross-Kerr Hamiltonian with single-qutrit rotations to form a controlled-SUM gate.
The foundation of our construction is the local equivalence between the controlled-SUM gate:
\begin{equation}
U_{\textrm{CSUM}}=\sum_{n=1}^{d}\dyad{n} \otimes X^{n}
\end{equation}
and the controlled-phase gate $U_{\textrm{C}\phi}$:
\begin{equation}
U_{\textrm{C}\phi}=\sum_{n=1}^{d}\dyad{n} \otimes Z^{n},
\end{equation}
which we will soon show is achievable via the cross-Kerr interaction.
Here 
\begin{equation}
    X = \sum_{j=1}^d \dyad{j+1}{j}, \,\,\,\,\,\, Z = \sum_{j=1}^d e^{i 2\pi j / d} \dyad{j}{j}
\end{equation} 
are the generators of the set of qudit Pauli operators.
The two gates are related via conjugation by a single-qutrit Hadamard $H$ gate:
\begin{equation}
    (I\otimes H^\dagger) U_{\textrm{C}\phi} (I\otimes H) =  U_{\textrm{CSUM}},
\end{equation}
which, as for qubits, interchanges the Pauli $X$ and $Z$ operators.
This realization also easily generalizes to slight modifications of the controlled-SUM gate: one can realize instead a controlled-MINUS gate by inverting the Hadamard, and one can interchange the control and the target qudits by switching which qudit is Hadamard conjugated. 

Specifying to qutrits, the controlled-phase gate $U_{\textrm{C}\phi}$ imparts a phase $2\pi/3$ to the $\ket{11}$, $\ket{22}$ states and the opposite phase $-2\pi/3$ to the $\ket{12}$, $\ket{21}$ states.
 
We have developed two methods for implementing this gate via the cross-Kerr interaction.  
The first, described in what follows, is conceptually simpler and uses only four single-qutrit pulses, but is prone to local dephasing and to errors arising from a nonzero cross-Kerr interaction with qutrits outside the target-control pair.
The second method addresses these errors via dynamical decoupling in the specific context of the five-qutrit teleportation protocol, and is described in conjunction with the protocol in Section~\ref{sec: qutrit scrambling}.

The essential concept behind these constructions is to intersperse time-evolution under the native cross-Kerr Hamiltonian with single-qutrit pulses exchanging the $\ket{1}$ and $\ket{2}$ states.
The simplest construction uses four segments of cross-Kerr evolution.

Denoting a swap pulse on qutrit $q$ as $\pi^{12}_{q}$, the total pulse sequence is given by the decomposition:
\begin{equation}
\begin{split}
     U_{C\phi} = & \,\, e^{-i T_A H_{cK}} \cdot \pi^{12}_0 \cdot e^{-i T_B H_{cK}} \\
     & \cdot \pi^{12}_1 \cdot e^{-i T_C H_{cK}} \cdot \pi^{12}_0 \cdot e^{-i T_D H_{cK}} \cdot \pi^{12}_1,
\end{split}
\end{equation}

where $T_A$, $T_B$, $T_C$, $T_D$ denote the time durations of each segment of cross-Kerr evolution. 
To determine these times from the interaction coefficients, note that each evolution segment serves to add a phase $\phi_{ij} = \alpha_{ij} T$ to the $\ket{ij}$ state (here, $i,j \in \{ 1,2\}$), while the swap pulses interchange the indices $1 \leftrightarrow 2$ on the acted upon qutrit.
Our choice of swap gates guarantees that each state spends exactly one segment under each interaction coefficient, and returns to itself at the end of the gate.
The unitary thus amounts to a phase applied to each state equal to a linear combination of the four evolution times, with the transformation matrix determined by the interaction coefficients.
For generic interaction coefficients this linear transformation from interaction times to applied phases is full rank,
which guarantees that the method can in principle generate \emph{any} combination of two-qutrit phases.

\begin{figure*}
\includegraphics[width=\textwidth]{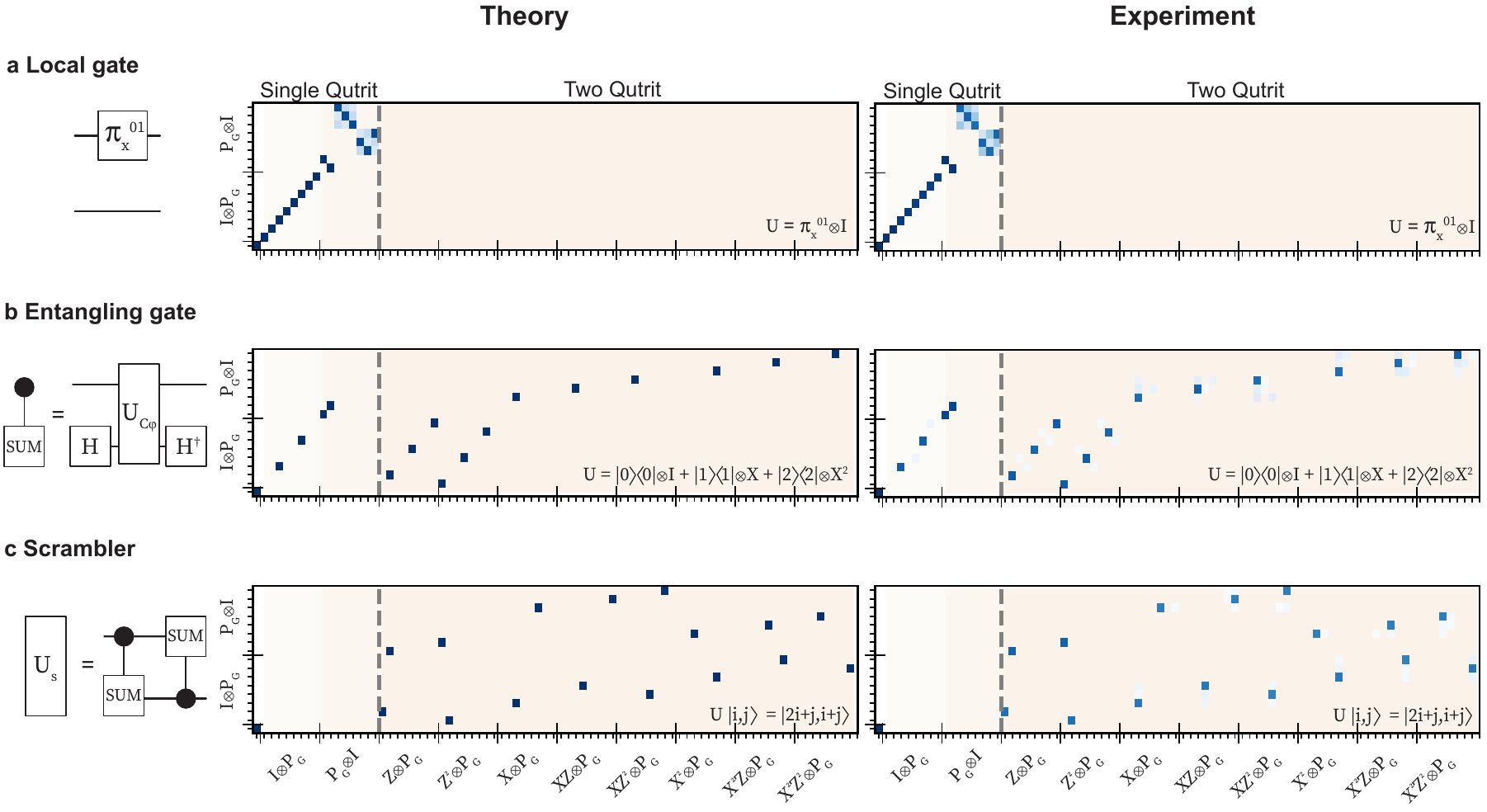}\caption{ \textbf{Quantum process tomography of the two-qutrit scrambling unitary} | Results of process tomography both experimental and ideal. Plotted is part of the Pauli transfer matrix (where the vertical axis only includes the single qutrit operators, and $P_G$ represents the single-qutrit Pauli group excluding $I$) which shows the unitary's action on single-qutrit Pauli operators when the unitary is a local gate (\textbf{a}), non-scrambling entangling gate (\textbf{b}) and the scrambling-unitary (\textbf{c}), .  This directly verifies the key characteristic of scrambling by $U_s$, that it maps all non-identity single-qutrit Pauli operators to two-qutrit operators.
\label{fig:QPT}}
\end{figure*}

With our particular interaction coefficients, this controlled-phase gate implementation takes a total time $\sim \! 1.5$ $\mu$s for each of the qutrit pairs $(Q_1,Q_2)$ and $(Q_3,Q_4)$. 
We characterize the performance of the full controlled-SUM gate via quantum process tomography on the full two-qutrit subspace (Fig.~\ref{fig:QPT}b) and achieve a fidelity of 0.889, primarily limited by decoherence occurring throughout the cross-Kerr time evolution.

Combined with the single-qutrit control demonstrated in the previous section, each of the conditional-$\pi$ gate and the conditional-SUM gate enable universal quantum computation on our qutrit processor~\cite{di2011elementary}. 
Having both gates on hand provides additional flexibility in gate compilation as the different entangling gates may be more adept in different scenarios.

\section*{III. Qutrit scrambling}\label{sec: qutrit scrambling}

We now turn to proof-of-principle experiments demonstrating the capabilities of our qutrit processor, specifically by demonstrating and verifying quantum scrambling.  We will demonstrate the implementation of a two-qutrit scrambling unitary, and verify that it scrambles using a five-qutrit quantum teleportation algorithm.  
Quantum scrambling, loosely defined as the delocalization of initially local operators induced by many-body time dynamics, is an area of active current research; it is likely that quantum simulations performed on future processors will deepen our understanding of this phenomenon and many-body quantum dynamics more generally.   To date, experimental realizations of scrambling have been performed with qubits; however, many interesting scrambling-related phenomena occur in higher-dimensional systems; see, e.g.,~\cite{gu2017local,nahum2017quantum,Zhuang_2019}.  Using quantum scrambling to demonstrate the capabilities of our processor thus also represents a step towards such simulations of these high-dimensional composite systems.

Specifically, we implement a simple Clifford scrambling unitary $U_{s}$, which effects the following permutation of the nine two-qutrit computational basis states~\cite{Yoshida_2019}: 
\begin{equation}
    U_{s} \ket{m,n} = \ket{2m+n, m+n}.
\end{equation}
As described in Ref.~\cite{Yoshida_2019}, the scrambling behavior of this unitary can be seen in the effect it has on the single-qutrit Pauli operators $X$ and $Z$: 
\begin{eqnarray}
    U\left(Z\otimes I\right) U^\dagger &=& Z \otimes Z^2 \\
    U\left(I\otimes Z\right) U^\dagger &=& Z^2 \otimes Z^2 \\
    U\left(X\otimes I\right) U^\dagger &=& X^2 \otimes X \\
    U\left(X\otimes I\right) U^\dagger &=& X \otimes X.
    \label{eq:scrambling_transformations}
\end{eqnarray}
Each single-qutrit Pauli operator is transformed into a two-qutrit operator, as required by scrambling.  Additionally, the transformations above show that $U_s$ is also a Clifford scrambler, i.e. a scrambling unitary which is a member of the Clifford group:  this is evidenced by the fact that the action of $U_s$ on Pauli operators is closed.  

For a bit of context, we note that scrambling is not possible in a two-qubit system: there is no unitary operation that completely delocalizes all single-qubit operators.  A system of two qutrits is the smallest bipartite system in which maximal scrambling is possible.  

In the next two sections, we verify the maximally-scrambling nature of our implementation in two ways: (i) explicitly through quantum process tomography, and (ii) through a five-qutrit teleportation protocol inspired by quantum gravitational physics.

\subsection*{A. Verifying scrambling through quantum process tomography}

On our processor, the scrambling unitary $U_s$ is constructed by applying two controlled-SUM gates in sequence, switching the control and target qubit between the two (Fig.~\ref{fig:QPT}c).   In order to fully characterize our implementation of the scrambling unitary $U_\textrm{s}$ we perform full quantum process tomography. 
For a two-qutrit unitary, this is a highly measurement-intensive procedure, which entails reconstructing, via two-qutrit state tomography, a $9 \times 9$ output density matrix for a complete set of 81 input states. 
This requires millions of measurements -- each involving state preparation, application of $U_\textrm{s}$, and state measurement (in a different basis) -- with the precise number of repetitions determined by the desired statistical uncertainty. 
For experimental platforms with a duty cycle in the Hz range (e.g. trapped ions) this procedure would be prohibitively long; in contrast, superconducting circuits can reach repetition rates up to $\sim 100$ kHz, such that the full process matrix can be measured in approximately 1 hour.  
This gives superconducting platform the unique ability to not only quantify the process fidelity of the implemented scrambling unitary but also to measure precisely \textit{how} it scrambles. 

We implement the scrambling unitary using two sequential controlled-SUM gates based on the cross-Kerr gate described in the previous section. 
In Fig.~\ref{fig:QPT}c, we depict the results of quantum process tomography on our implementation. 
Through this tomography, we find that the fidelity of the scrambling operation on our hardware is 0.875, with two dominant error mechanisms: (i) dephasing and (ii) amplitude-damping during the cross-Kerr evolution.

Quantum process tomography also allows us to directly visualize the entire action of the unitary via its effect on local operators; i.e., we can directly visualize (Fig.~\ref{fig:QPT}) the relations (13)-(16). 
In particular, in Fig.~\ref{fig:QPT}c we can verify explicitly that $U_{s}$ transforms all single-qutrit operators into fully two-qutrit operators -- the definition of a maximally scrambling unitary.
We can also verify that it does so as a Clifford unitary, i.e. all Pauli operators are transformed into different Pauli operators.
For comparison, we also illustrate the quantum process maps, both theoretical and experimental, of a single-qutrit unitary that does not delocalize any information (Fig.~\ref{fig:QPT}a), as well as a controlled-SUM gate, which is entangling but not fully scrambling (Fig.~\ref{fig:QPT}b).  
In the former, we verify that no single-qutrit operators are delocalized, while in the latter we observe that all single-qutrit operators are delocalized except for $Z^{(\dagger)} \otimes I$ and $I \otimes X^{(\dagger)}$, which commute with the controlled-SUM gate.

\subsection*{B. Verifying scrambling through quantum teleportation}

\begin{figure} 
\includegraphics[scale=1]{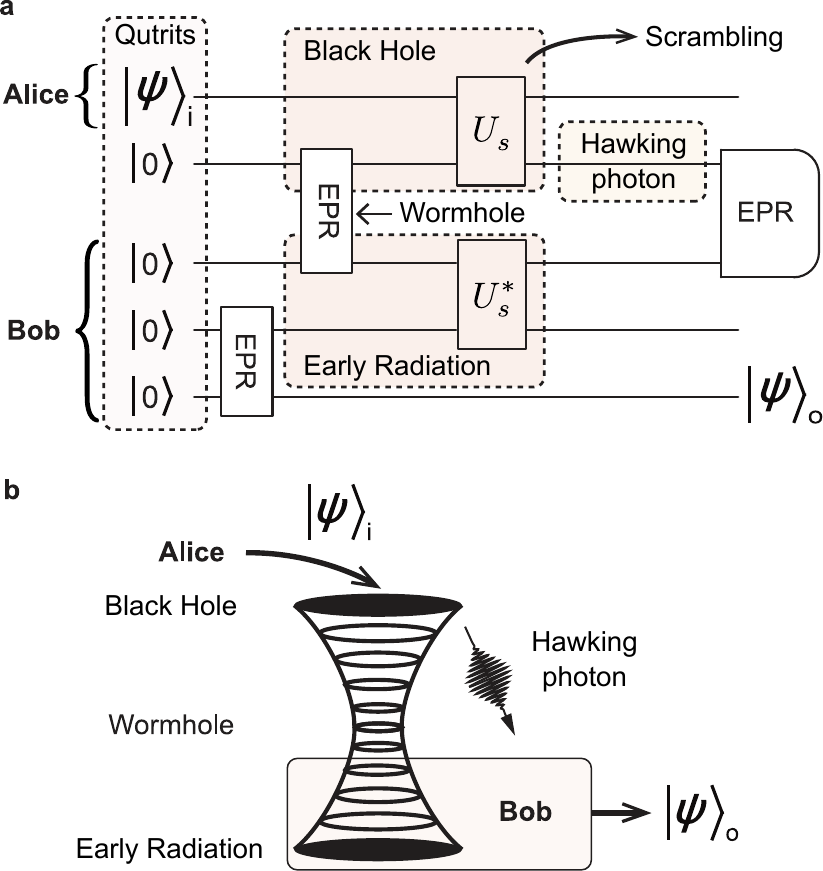}
\caption{ \textbf{Quantum teleportation circuit | }
Five-qutrit teleportation protocol used to test for scrambling by the two-qutrit unitary $U_{s}$, and its interpretation in the context of black hole physics.  
\textbf{a,} We begin with the first qutrit in a quantum state $\ket{\psi}$, and prepare the remaining qutrits into two EPR pairs.
The first qutrit $Q_1$ is `scrambled' with half of the first EPR pair by the unitary $U$, while the conjugate $U^*$ is applied to $Q_3$ and $Q_4$.
An EPR measurement on a pair of qutrits ($Q_2$ and $Q_3$) exiting each of $U$, $U^*$ serves to teleport the initial quantum state to $Q_5$ if and only if $U$ is scrambling.
\textbf{b,} In the context of the Hayden-Preskill thought experiment, $Q_1$ corresponds to Alice's diary, which subsequently falls into a black hole ($Q_2$) that is maximally entangled with its past Hawking radiation ($Q_3$),  held by an outside observer Bob.
In the gravitational picture, this shared entanglement functions similarly to a geometric wormhole between the black hole and its early radiation.
The black hole dynamics are modelled as a scrambling unitary $U$, followed by emission of a Hawking photon ($Q_2$).
Finally, Bob applies $U^*$ and measures in the EPR basis to recover Alice's diary.}
\label{fig:fig1}
\end{figure}

While  process tomography provides an elegant and exhaustive way to ``image'' scrambling performed by our two-qutrit unitary, such an approach is infeasible for verifying scrambling in larger systems, as the number of required measurements scales exponentially with system size. 
As an example of a protocol which can circumvent this difficulty, and thus could be used on larger quantum processors, we turn to a teleportation protocol that quantifies the scrambling behavior of a unitary by using scrambling to enable the teleportation of quantum states~\cite{yoshida2017efficient,Yoshida_2019, Landsman_2019}.
Compared to tomography, a downside of the teleportation protocol is that it is limited as an essentially `one-parameter' diagnostic of scrambling: it quantifies the average amount that operators on a given qutrit are transformed to have support on an additional qutrit.  
It does so via an average of OTOCs associated with the unitary over local operators; in this context, maximal scrambling by $U_{s}$ is captured by the fact that the average OTOC decays to its minimum allowed value ($1/9$ for a two-qutrit system)~\cite{Yoshida_2019}. 
Crucially, the protocol is constructed in such a way that faithful teleportation of a quantum state $\ket{\psi}$ \textit{requires} quantum information scrambling to occur, as well as the absence of experimental errors. 
The teleportation fidelity can then in turn be used to upper-bound the OTOC, even in the face of such errors.  

As shown in Fig.~\ref{fig:fig1}, the verification protocol requires both the scrambling unitary, $U_{s}$, and its time-reversal, $U_{s}^*$, to be performed in parallel on separate pairs of qutrits.  The qutrit pairs undergoing these two time-evolutions are initially highly correlated: two qutrits out of the four (one from each pair) begin as a maximally entangled EPR pair. After applying $U_{s}, U_{s}^*$, the probability of measuring this same pair in the EPR state decreases to $1/9$ -- the same value as in a maximally mixed state -- due to scrambling with the additional qutrits.
However, simply observing this decrease is not enough to robustly demonstrate scrambling, since the same decrease could also have resulted from deterioration of the EPR pair under experimental decoherence. 

Here, teleportation comes to our aid: we place one of the remaining two qutrits ($Q_1$, the `input') in an arbitrary pure single-qutrit state $\ket{\psi}$, and the other ($Q_4$) in an EPR pair with a fifth qutrit ($Q_5$, the `output').  
In the absence of experimental error, maximally scrambling dynamics guarantee that whenever $Q_2$ and $Q_3$ happen to be measured in their initial EPR state, the state $\ket{\psi}$ of $Q_1$ is teleported to $Q_5$.  
Unlike measuring a low EPR probability, high-fidelity teleportation can only arise from quantum information scrambling, not as a result of decoherence, making the teleportation fidelity a robust diagnostic of information scrambling \cite{yoshida2017efficient,Yoshida_2019}.  
Furthermore, while experimental error will lead to a decay of the teleportation fidelity from unity, any measured fidelity above the classical limit (0.5 for qutrits) places a non-trivial upper-bound on the averaged OTOCs, and thus the scrambling behavior, of $U_{s}$~\cite{Yoshida_2019}.

This association between scrambling and teleportation originated in black hole physics, and the teleportation protocol explored here  is in fact inherited directly from this context.
Its most straightforward interpretation is based on the Hayden-Preskill variant of the black hole information paradox~\cite{hayden2007black,yoshida2017efficient}, as outlined in Fig. \ref{fig:fig1}.
Here, one pictures an observer Alice who has dropped a `diary' consisting of the quantum state $\ket{\psi}$ into a black hole.
Meanwhile, an outside observer Bob wonders whether this information is recoverable, as opposed to being irreversibly destroyed by the black hole.
In seminal works~\cite{hayden2007black,yoshida2017efficient}, it was shown that \textit{if} ($i$) the black hole's dynamics are approximated as a fully scrambling quantum unitary $U_s$ and ($ii$) Bob possesses a large resource of entanglement with the black hole (e.g. a collection of its early Hawking radiation), then the state $\ket{\psi}$ can in fact be recovered via quantum operations on any few degrees of freedom emitted from the black hole (e.g. one additional Hawking photon).

In the gravitational picture, the shared entanglement between the black hole and its early radiation functions similarly to a geometric wormhole between the two systems.
Indeed, in a close variant of this protocol, teleportation has a precise interpretation as Alice's state traversing a two-sided wormhole, in the specific case where the unitary corresponds to time-evolution under many-body dynamics with a holographic gravity dual~\cite{Gao_2017,maldacena2017diving,brown2019quantum}.
Quite surprisingly, recent work has revealed that the \textit{same} protocol also features a more generic, non-gravitational mechanism for teleportation, which displays remarkably similar features to gravitational teleportation but is based only on the spreading of operators~\cite{yoshida2017efficient,schuster2020teleportation}.
This second form of teleportation encapsulates the physics observed here.

\begin{figure*}
\includegraphics[scale=1]{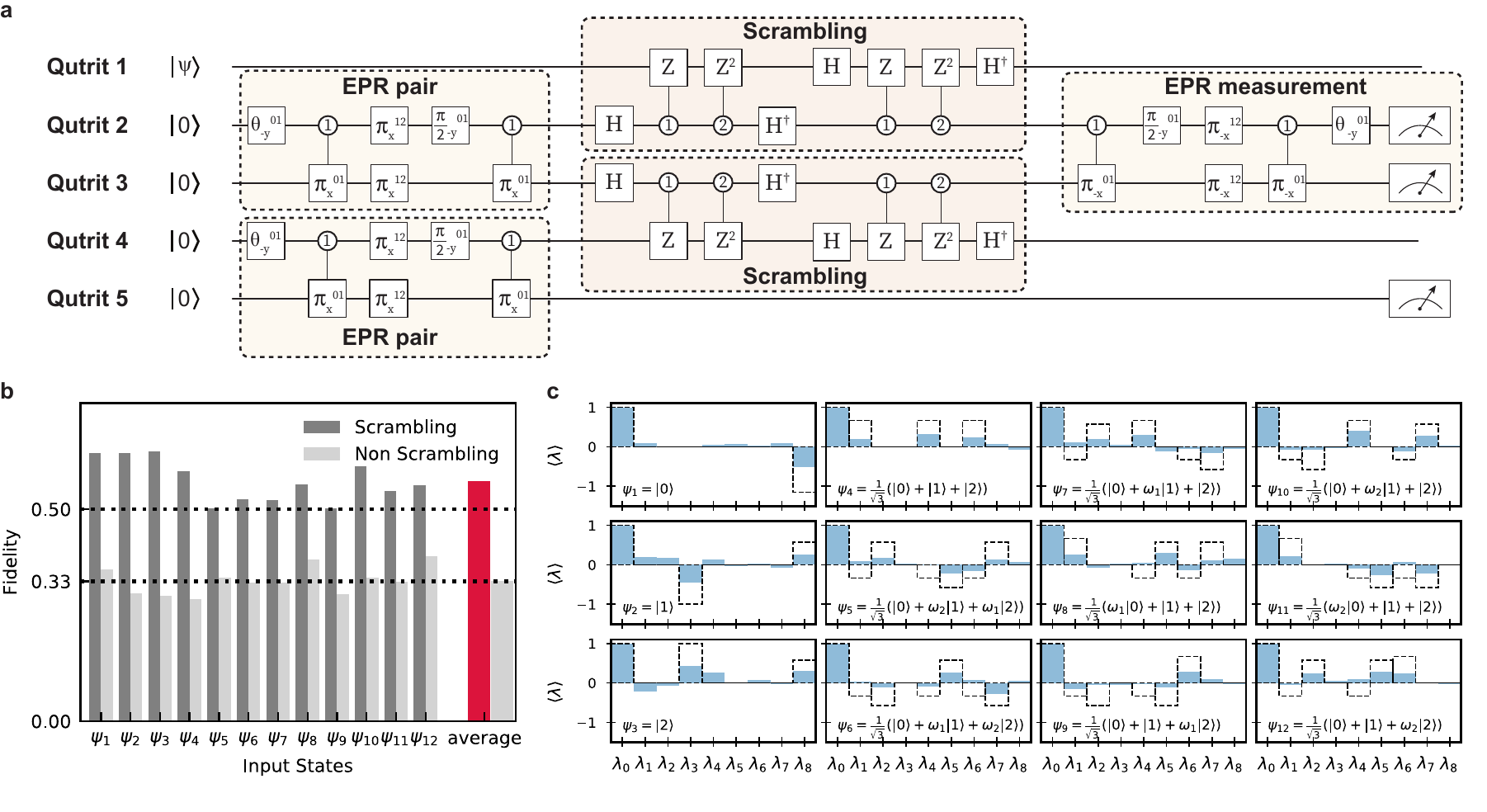}\caption{ \textbf{Results of the five-qutrit teleportation protocol | a,} An expanded view of the five-qutrit teleportation protocol in Fig. \ref{fig:fig1}, showing the native operations used to realize each portion of the protocol. \textbf{b,} Measured teleportation fidelities for twelve teleported states, which combine to give an unbiased estimate of the teleportation fidelity averaged over \textit{all} single-qutrit pure states.  An average fidelity above $1/2$ -- the classical limit for teleportation of qutrits -- verifies non-zero quantum information scrambling by the maximally scrambling unitary, despite the presence of experimental error. When the scrambling unitary is replaced with an identity operation with same complexity as the scrambler, the average teleportation state fidelity reduces to $1/3$, the same as a random guess when one does not have access to the input state.  \textbf{c,} Representation of each of the twelve reconstructed density matrices after teleportation, expressed in the basis of Gell-Mann matrices ($\lambda_0 -\lambda_8$) with dotted lines showing the ideal result.
\label{fig:Teleportation}}
\end{figure*}

Experimentally, the teleportation protocol requires three essential ingredients: initial preparation of EPR pairs, implementation of $U_{s}$ and $U_{s}^*$, and measurement in the EPR basis.
We have discussed our implementation of the first two of these ingredients in previous sections. As our architecture only natively allows for computational-basis measurements, we  realize the EPR measurement on $Q_2$ and $Q_3$ by performing the EPR creation sequence in reverse, which transforms $\ket{\textrm{EPR}}$ to the computational basis state $\ket{00}$. We then measure all five qutrits in the computational basis and use the measurements outcomes showing $Q_2$ and $Q_3$ in $\ket{00}$ as a herald for successful teleportation. We verify success by tomographically reconstructing the quantum state of $Q_5$ after EPR measurement (by inserting state tomography pulses on $Q_5$ before the final measurement), represented by the density matrix $\rho_{\text{out}}$.  

All three of the steps described above work particularly well on pairs of qutrits in ``isolation'', i.e.~when the pair's neighboring qutrits are left in their ground states. 
However, this is necessarily \textit{not} the setting of any five-qutrit quantum circuit, including the teleportation protocol. Here, we have to contend with the ``always-on'' nature of the cross-Kerr interaction, which leads to unwanted entangling dynamics between all neighboring qutrit pairs, even as we attempt to perform our quantum gates. To combat these unwanted interactions, we develop a novel set of dynamical decoupling sequences optimized for qutrits (see Appendix \ref{app:DD} for details). 

We perform teleportation of twelve different single-qutrit quantum states $\ket{\psi}$, corresponding to all possible eigenstates of single-qutrit Pauli operators, i.e. three eigenstates each for $X, Z, X Z, X Z^2$ (the remaining Pauli operators are either inverses of these or the identity). We calculate the teleportation fidelity from the reconstructed density matrix of the output qubit as $F_\psi = \bra{\psi} \rho_{\text{out}} \ket{\psi}$.
Our chosen set of states is known as a `state 2-design', which guarantees that the average teleportation fidelity $F_{\text{avg}} = (1/12) \sum_\psi F_\psi$ measured for these twelve sampled states is in fact equal to the average teleportation fidelity over \textit{all} possible states~\cite{Roberts_2017,Yoshida_2019}.
Furthermore, this average teleportation fidelity in fact allows us to upper bound the average OTOCs associated with the implemented unitary by $(4F-1)^{-2}$, without making \textit{any} assumptions about the nature of the noise affecting our quantum processor~\cite{Yoshida_2019}.

The results of teleportation for two different choices of unitary dynamics are shown in Fig.~\ref{fig:Teleportation}.
First, as a control, we perform the protocol with the identity operation in place of $U_s$. 
To ensure that the two have similar magnitudes and modes of experimental error, we implement the identity with precisely the same number and type of gates as the maximally scrambling unitary;  the two differ only in the magnitude of certain software-defined phase gates. 
Since the identity operator does not scramble quantum information, we observe trivial teleportation fidelities $F_\psi \approx 1/3$ for all input states (Fig.~\ref{fig:Teleportation}b).
Indeed, using quantum state tomography we verify that the final state of qubit $Q_5$ is near maximally mixed regardless of input state.

Finally, we perform the teleportation protocol with the maximally scrambling unitary $U_s$, which in theory should allow for perfect teleportation of any input quantum state.
In contrast to the identity operation, we observe that all but one of the input states are teleported with fidelity $F_\psi > 1/2$, with an average fidelity of $F_{\textrm{avg}} = 0.568 \pm 0.001$.   
This allows us to experimentally upper bound the averaged OTOC by $0.618 \pm 0.004$.
This marks a significant decay from its non-scrambling value of unity and verifies that our platform is capable of coherently simulating multi-qutrit scrambling dynamics. 

\section*{IV. Conclusion}
In summary, we have demonstrated a five-qutrit quantum processor built from superconducting transmon circuits. 
Our qutrit toolbox features high fidelity single- and two-qutrit gates, as well as characterization methods such as state- and process-tomography that provide useful information for benchmarking and debugging large-scale computations.
Using the verification of qutrit information scrambling as a proof-of-principle task, we have demonstrated two distinct entangling gates as well as a simple dynamical-decoupling protocol that allows the gates to be run simultaneously on adjacent qutrits.  
Interestingly, this experiment can be also interpreted as an implementation of quantum error-correction. 
In this language, the teleportation protocol is equivalent to a three-qutrit quantum error-correcting code, which protects information from the erasure of any one of the three qutrits \cite{cleve1999share}.  Exploring other forms of error-correction in qudit-based processors is a natural extension of our work.

A number of intriguing future directions are suggested by this work.  First, our platform opens the door to exploring the potential advantages of ternary quantum logic, including a more efficient decomposition of the Toffoli gate~\cite{gokhale_arxiv_2019} as well as magic state distillation protocols that outperform conventional qubit-based strategies in terms of both error-threshold and yield~\cite{campbell2012magic}. Second, the dynamical decoupling techniques introduced here naturally apply to other spin-1 systems including solid-state defect centers and multi-level atomic, molecular and trapped ion systems~\cite{senko2015realization, choi_PRL_2017}. Third, one can imagine further enlarging the qudit dimension by further leveraging the intrinsic anharmonicity of transmons~\cite{gao_Nature_2019,naik_NatComm_2017}, enabling the study of many-body phases and entanglement  dynamics in higher-spin quantum systems~\cite{pai2019localization}. Finally, building upon recent excitement surrounding quantum supremacy protocols using pseudorandom quantum circuit sampling~\cite{arute2019quantum}, it would be natural to investigate analogous qudit-based protocols, where supremacy might be achieved using a substantially smaller number of elements. \\

\begin{acknowledgments}
\noindent \textbf{Acknowledgements} The authors gratefully acknowledge the conversations and insights of James Colless, Emmanuel Flurin, Daniel Harlow, William Livingston, Leigh S. Martin, Brad Mitchell and Brian Swingle. This work was supported by multiple grants awarded by the Department of Energy. From the Office of Advanced Scientific Computing Research: the Advanced Quantum Testbed and the Quantum Algorithm Teams Programs. From the Office of High Energy Physics: the HEP QUANTISED and the COMPHEP pilot ``Probing information scrambling via
quantum teleportation'' (DE-AC02-05CH11231).
T.S. acknowledges support from the National Science Foundation Graduate Research Fellowship Program under Grant No. DGE 1752814. \\
\end{acknowledgments}

\appendix

\section{Processor and fabrication details}
\label{app:Fab}
The processor we use features five fixed-frequency (single junction) transmon qutrits on a chip with an eight-transmon ring geometry.  The readout and coupling resonators, Purcell filter, transmon capacitors, microwave drive lines and ground plane are composed of niobium, while the transmon junctions are aluminum with an aluminum oxide barrier (Fig 1a main text).  

The processor is fabricated on intrinsic $>$8000 ohm-cm silicon $<$100$>$ wafers.  Initial cleaning of the silicon wafer occurs in piranha solution---a mixture of sulfuric acid and hydrogen peroxide---at 120$^{\circ}$C for 10 minutes, followed by 5:1 buffered oxide etch (BOE) for 30 seconds to remove surface contaminants and native oxides.  A 200-nm thick film of niobium is then sputtered onto the wafer, with deposition pressures optimized to yield a slightly compressive film.  Following this, junctions and all other structures are patterned using 3 rounds of electron-beam lithography.  We use MicroChem MMA EL-13 copolymer as a resist, developing it in a 3:1 mixture of IPA:MIBK (isopropyl alcohol and methyl isobutyl ketone) at room temperature for 8 minutes.  We then etch the niobium with chlorine chemistry in an inductively coupled reactive ion etcher with about 50 nm overetch into the silicon. After etching, resist is removed with Microposit 1165 at 80$^{\circ}$C for 60 min.  The Josephson junction fabrication process begins by stripping native Nb and Si oxides with 30 seconds in BOE.  Resist is then spun: we use 500 nm of MMA EL-13 and 150 nm of AllResist AR-P 6200.9, both baked at 150$^\circ$C for 60 and 90 seconds, respectively. We write "Manhattan style" junction patterns \cite{Costache2012, Potts2001} (proximity-effect-corrected with Beamer by Genisys software) at 100 keV in a Raith EBPG 5150 using a 200 pA beam current and 200 $\mu$m aperture. After writing, the exposed AR-P is first developed in n-amyl acetate chilled to $0^\circ$C; after this the AR-P development is halted with 10s immersion in IPA; finally MMA is developed in 3:1 IPA:MIBK for 10 min.  We then dry the resulting structure with $N_2$ and descum it with an 80W, 200 mbar oxygen plasma etch.  This etching step is split into 4 separate substeps, with 90 degree substrate rotations between each substep for improved junction uniformity. Newly-formed oxides at the bottom of the developed structure are then removed with a 15s dip in BOE.  The wafer is then loaded into a Plassys MEB550s evaporator and pumped overnight before the junction evaporation steps: first, an Al base electrode is evaporated and the tunnel barrier then formed by thermal oxidiation, introducing a $95 \% / 5\%$ Ar/O mix into the chamber at 10 mbar for 10 min. A second aluminum electrode is then evaporated to complete the junction and a third evaporation is necessary to climb the second 250 nm capacitor step edge. The junction pattern includes a 6 x 8 $\mu$m Al wire on top of the Nb for electrical contact between the junction and capacitor. After liftoff for 2 hours in acetone at 67$^\circ$C, the same resist stack is spun, and 10 x 15 $\mu$m rectangles are opened over the Al/Nb overlap region. The exposed metals are then ion milled and Al is subsequently e-beam evaporated to ensure a low loss galvanic connection between Nb and Al~\cite{Dunsworth_2017}. More details on junction fabrication, including the steps leading to higher uniformity, can be found in~\cite{Kreikebaum2019}. After fabrication, the wafer is diced into 1x1 cm dies; cleaned in Microposit 1165 for 12 hours at 80C; sonicated in DI water, acetone, and IPA; descummed in 100 W oxygen plasma for 1 min and then wirebonded into a gold plated copper cryopackage on a 300 $\mu$m air gap.

Each transmon is coupled to ($i$) a linear readout resonator to enable multiplexed dispersive measurement, ($ii$) two coupling resonators to enable entangling interactions with nearest neighbors, and ($iii$) a microwave drive line.  Readout resonators are loaded so that their effective linewidth $\kappa_{\textrm{ext}}\approx1$ MHz.  All readout resonators on the chip are coupled to a common $\lambda/2$ resonator, a Purcell filter with an external Q $\approx$ 10~\cite{Sete2015}.  The Purcell filter's passband overlaps with all readout resonator frequencies, allowing fast readout; all qutrit frequencies lie outside the passband, suppressing qutrit relaxation through this channel.  Slotline modes of all structures are suppressed using wirebonds; a wirebond also enables the readout bus to overlap a coupling resonator.  
\nocite{*}

\section{Experimental setup}
\begin{figure*}
\includegraphics[scale=1]{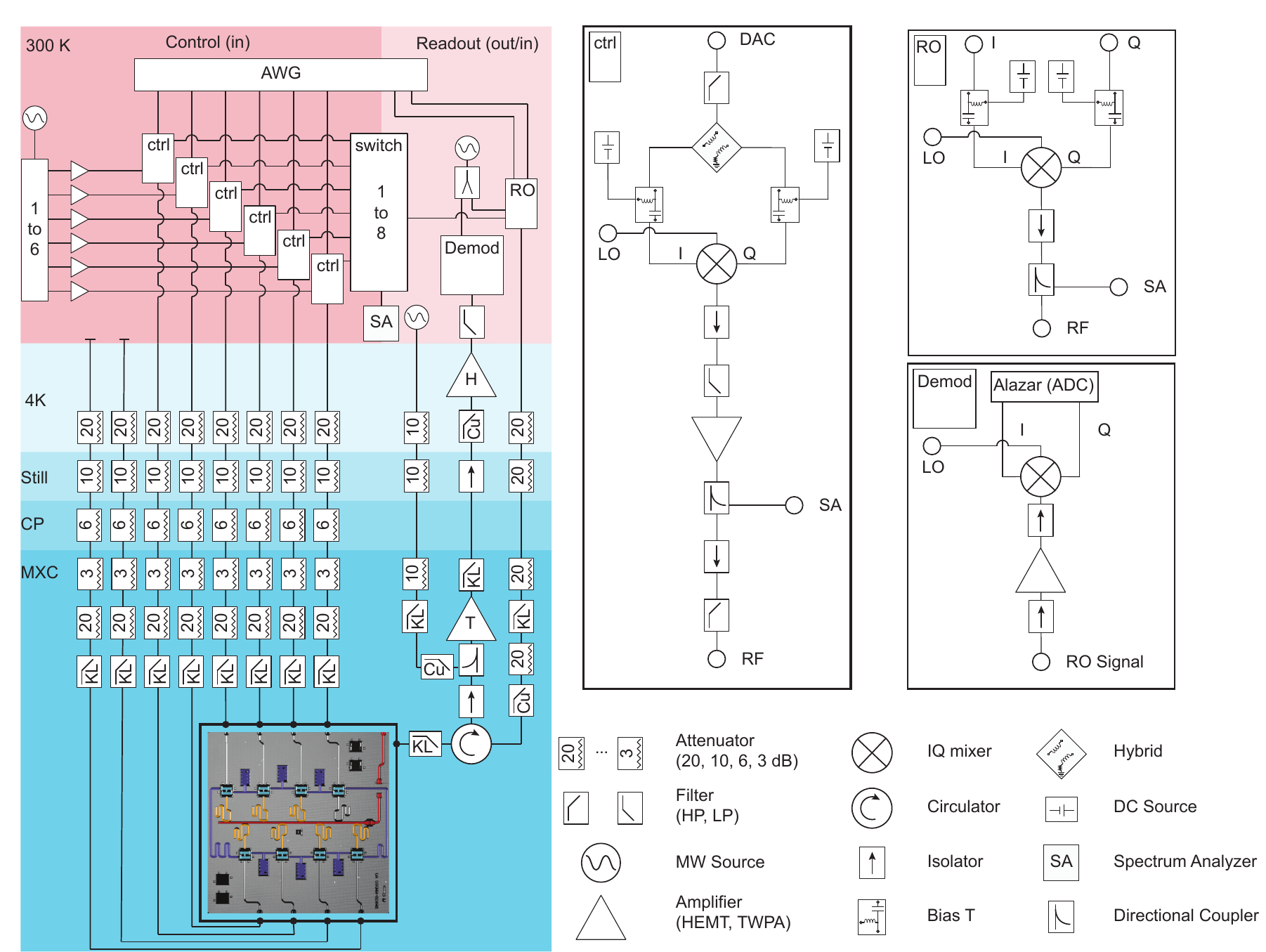}
\caption{ \textbf{Experimental setup}      
\label{fig:expsetup} described in detail in the text.}
\end{figure*}

The processor is installed in the 10 mK stage of a BlueFors XLD dilution refrigerator.  Room-temperature and cryogenic electronics for performing control and measurement of the qutrit chip are shown in Fig.~\ref{fig:expsetup}.   A Holzworth multi-channel synthesizer generates three local oscillator tones: a qubit control LO at 4.72 GHz, a readout LO at 6.483 GHz, and a pump at 7.618 GHz for the traveling-wave parametric amplifier (TWPAs).  Qutrit control pulses are formed by IQ modulating the amplified qubit LO (split six ways) with IF signals from a Tektronix AWG (sample rate 2.5~GS/s) with frequencies between 0.5 and 1.1 GHz.  We use both single-channel, hybrid-enabled SSB modulation and high-pass filtering to eliminate the lower sideband of the pulse, with additional band-pass filtering at room-temperature to eliminate noise from the AWG itself.  Readout signals are generated using two-channel SSB modulation with IF signals from the same Tektronix AWG.  All input signals are further attenuated in the cryostat.  

Readout signals are amplified by the TWPA at 10 mK, high-electron mobility transistor amplifiers (HEMT) at 4K, and further amplification at room-temperature before being digitized at 1.25 GSa/s and demodulated in software.

\subsection{Chip characterization}

Transmon parameters are given in Table~\ref{Table:chip_properties}.  The frequencies are extracted using standard spectroscopy methods.  Lifetimes are extracted by fitting decay curves to a single model with five parameters: two energy-relaxation times ($T_1^{1\rightarrow0}$ and $T_1^{2\rightarrow1}$) and a dephasing time for each basis state.  We perform randomized benchmarking to measure pulse errors of single qubit operations in the different subspaces, shown in the table.  

\begin{table*}
\centering
 \begin{tabular}{l  c  c  c  c  c} 
 \hline
  & $Q_1$ & $Q_2$ & $Q_3$ & $Q_4$ & $Q_5$ \\ 
 \hhline{======}
 Qutrit $\ket{0}\leftrightarrow\ket{1}$ frequency, $\omega_{01}/2\pi$ (GHz)~~~~ & 5.447 & 5.634& 5.776 & 5.619 & 5.431\\ 
 \hline
  Qutrit $\ket{1}\leftrightarrow\ket{2}$ frequency, $\omega_{12}/2\pi$ (GHz) &5.177 & 5.368 & 5.512 & 5.351 & 5.160 \\
 \hline
 Readout frequency, $\omega_{\textrm{RO}}/2\pi$ (GHz)  &6.384& 6.324 & 6.731 & 6.673 & 6.618 \\
 \hline
 Lifetime $T_1^{\ket{1}\rightarrow\ket{0}}$ ($\mu$s) & 70 & 49 & 43 & 55 & 63 \\ 
 \hline
 Lifetime $T_1^{\ket{2}\rightarrow\ket{1}}$ ($\mu$s) & 38 & 29 & 39 & 32 & 36 \\ 
 \hline
 Ramsey decay time $T_2^*$, $\ket{1}$/$\ket{0}$ ($\mu$s) & 73 & 13 & 41 & 48 & 20\\ 
 \hline
 Ramsey decay time $T_2^*$, $\ket{2}$/$\ket{1}$ ($\mu$s) & 13 & 10 & 16 & 23 & 10\\ 
 \hline
 Ramsey decay time $T_2^*$, $\ket{2}$/$\ket{0}$ ($\mu$s) & 16& 6 & 15 & 26 & 11\\ 
 \hline
 Echo time $T_{2\textrm{Echo}}$, $\ket{1}$/$\ket{0}$ ($\mu$s) & 71 & 51 & 46 & 64 & 74\\ 
 \hline
 Echo time $T_{2\textrm{Echo}}$, $\ket{2}$/$\ket{1}$ ($\mu$s) & 29 & 22 & 22 & 35 & 32\\ 
 \hline
 Echo time $T_{2\textrm{Echo}}$, $\ket{2}$/$\ket{0}$ ($\mu$s) & 39 & 26 & 34 & 45 & 39\\ 
 \hline
 Readout fidelity, $\ket{0}$ & 0.99 & 0.99 & 0.97 & 0.98 & 0.99 \\ 
 \hline
  Readout fidelity, $\ket{1}$ & 0.97 & 0.95 & 0.94 & 0.95 &0.96 \\ 
 \hline
  Readout fidelity, $\ket{2}$ & 0.95 & 0.94 & 0.92 & 0.95 & 0.96 \\ 
 \hline
 Per-Clifford error, $\ket{1}$/$\ket{0}$ subspace~~~~~ & 3.6e-4 & 3.9e-4  & 5.5e-4 & 2.7e-4 & -\\
 \hline 
 Per-Clifford error, $\ket{2}$/$\ket{1}$ subspace & 3.6e-4 & 6.0e-4 & 5.0e-4 & 7.5e-4 & - \\
 \hline

\end{tabular}
 \caption{Measured properties of the five qutrits} 
 \label{Table:chip_properties} 
\end{table*}

We further measure the coefficients of the cross-Kerr (or `ZZ') interaction by performing a Ramsey measurement with neighboring qutrits in the $\ket{1}$ or $\ket{2}$ states.  The cross-Kerr Hamiltonian between neighboring qutrits is 

\begin{eqnarray}
H_{\mathrm{Kerr}} / \hbar=\alpha_{11}|11\rangle \langle 11|+\alpha_{12}| 12\rangle\langle 12| \nonumber\\
+\alpha_{21}| 21\rangle\langle 21|+\alpha_{22}| 22\rangle\langle 22|
\label{eq:SuppCrossKerr}
\end{eqnarray}

Table \ref{Table:interactions} in the main text gives the value of these coefficients.  Residual cross-Kerr interaction coefficients between non-nearest-neighbor transmons were found to be negligible.  

The cross-Kerr interaction is the dispersive limit of an exchange interaction between nearest-neighbor transmons mediated by the coupling resonators, governed by the Hamiltonian
\begin{equation}
H_{\textrm{int}}/\hbar=g\left(a^{\dagger} b+b^{\dagger} a\right)
\end{equation}
We measure the value of $g$ on a chip with tunable transmons, but otherwise identical to the one used in the experiment.  Spectroscopy of the avoided crossing reveals an interaction amplitude $g$ of roughly 3 MHz.  

\subsection{Coherence of third transmon level}

The dominant noise processes affecting transmons tend to worsen for states higher up the transmon ladder. For our qutrit-based processor there are two salient manifestations of this: 
\begin{itemize}
    \item Due to bosonic enhancement, amplitude damping (spontaneous emission) decay from state $\ket{2}$ to $\ket{1}$ proceeds roughly twice as fast as the decay from $\ket{1}$ to $\ket{0}$.  
    \item Dephasing due to charge noise, which randomizes the relative phases between the $\ket{0}$, $\ket{1}$, and $\ket{2}$ states, occurs roughly an order of magnitude faster for each state up the transmon ladder:  in particular, for the $\ket{2}$ state relative to the $\ket{1}$ state.   
\end{itemize} 
As stated in the main text, careful fabrication, microwave engineering, and parameter selection were required to obtain high coherence in the transmon qutrit. The fabrication and microwave engineering are detailed in Appendix \ref{app:Fab}, and served to mitigate the $T_1$ decay.  Here we describe the parameter selection -- specifically, the choice of the transmon $E_J/E_C$ ratio -- which was chosen to minimize the effect of dephasing.

Transmons are characterized by two parameters: the Josephson energy $E_J$, and the capacitive energy $E_C$~\cite{Koch_2007}.  Increasing the $E_J/E_C$ ratio exponentially decreases the sensitivity of all transmon eigenstates to charge noise, at the expense of also lowering the transmon's anharmonicity.  Specifically, the charge dispersion $\epsilon_m$ of the $m^{\text{th}}$ level is given by
\begin{equation}
    \epsilon_m \approx (-1)^m E_C \frac{2^{4m+5}}{m!}\sqrt{\frac{2}{\pi}}\left(\frac{E_J}{2E_C} \right)^{\frac{m}{2}+\frac{3}{4}} e^{-\sqrt{8E_J/E_C}},
\end{equation}
while the relative anharmonicity $\alpha_r$ is given by
\begin{equation}
    \alpha_r \approx - (8E_J/E_C)^{-1/2}.
\end{equation}
Typical transmon qubit designs use ratios $E_J/E_C \approx 50$.
We initially used such a ratio, which resulted in charge dispersion of 102 kHz and $<$10 kHz of $\ket{2}$ and $\ket{1}$ states, respectively.  However, with these parameters, charge-parity fluctuations~\cite{Riste_2013} dephase the coherence between the $\ket{2}$ and $\ket{1}$ states within 5 $\mu$s, making high-fidelity gates impossible to implement.  To mitigate this dephasing, we switched to a design with $E_J/E_C \approx 73$, which resulted in charge dispersions of 12 kHz and 261 Hz for the $\ket{2}$ and $\ket{1}$ states, respectively.  This also reduced the anharmonicity from roughly 300 MHz to roughly 250 MHz.

\subsection{Crosstalk}
As discussed in the main text, each transmon features a dedicated microwave control line through which we drive single- and two-qutrit gates.  However, we find significant (order unity compared to intended coupling) crosstalk between the microwave drive lines for each qutrit.  This crosstalk is non-local, not confined to nearest or next-nearest neighbors.  When driving Rabi oscillations on a given qutrit, it produces two unwanted effects:
\begin{enumerate}
    \item All other qutrits will be off-resonantly Rabi driven.  Depending on the relative frequencies between the qutrits, this can either manifest as an unwanted change in a qutrit's state populations (if the frequencies are relatively close) or a Stark shift (if the frequency difference is large compared to the Rabi frequency).  
    \item Microwave field leaking onto one or more neighboring qutrits will result in an unwanted cross-resonance interaction, making the desired Rabi oscillation frequency vary with the state of the neighboring qutrit(s).  This effect was anticipated in ~\cite{RigettiDevoret}.
\end{enumerate}

We observed no indications of nonlinearity in the cross-talk at the drive powers we used.  That is, for a given drive frequency, the cross-talk can be characterized in terms of a five-by-five complex-valued matrix $C(\omega)$ relating the field amplitudes $\vec{o}(\omega)$ seen by each of the five qutrits to the input field amplitudes $\vec{i}(\omega)$ on each drive line: $\vec{o}(\omega)=C(\omega) \vec{i}(\omega).$ We did observe a strong frequency-dependence of the cross-talk matrix. 

The linearity of the cross-talk enabled us to compensate for it by inverting the matrix $C(\omega)$ at each drive frequency, yielding combinations of microwave drive lines which would route the drive field to only a single qutrit.   The main challenge in this scheme was the measurement of $C(\omega)$.  Our strategy was to focus on two drive lines at a time, and find for each line the relative amplitudes and phases which exactly cancelled the field at the location of all of the qutrits on our chip---depending on the relative frequencies, we used either a Stark shift or a Rabi oscillation as a symptom of unwanted microwave field.  This measurement was repeated for each of ten drive frequencies of interest (i.e. the $|0\rangle \leftrightarrow|1\rangle $ and $|1\rangle \leftrightarrow|2\rangle$ transition frequencies of all five qutrits), each pair of lines, and each qutrit on the chip.   

Our crosstalk cancellation method is extremely measurement-intensive and was feasible only because of the relatively few qutrits in this work.  On future quantum processors with tens or hundreds of quantum systems, the number of measurements required for our cancellation scheme would be prohibitively expensive.  In addition, the strong frequency dependence of the cross-talk matrix limits the speed at which one can apply single-qudit pulses in this manner:  for pulses approximately 10 ns in length, we observed cross-talk which we could not compensate for using our method, likely because of this frequency dependence combined with Fourier broadening of the pulses.  Going forward, it is thus important to pinpoint the source of the microwave cross-talk, in order to develop scalable solutions at the hardware level.

\section{Qutrit Operations and gate-set}
\label{app:QutritGates}
\subsection{Single Qutrit Operations}

A convenient set of generators to describe qutrit rotations are the Gell-Mann matrices:
\begin{widetext}

\begin{align*}
  \lambda_1 \equiv s_x^{01}& =  \matthree {0}{1}{0}{1}{0}{0}{0}{0}{0} &
  \lambda_2 \equiv s_y^{01}&= \matthree {0}{-i}{0}{i}{0}{0}{0}{0}{0}&
  \lambda_3 \equiv s_z^{01}&= \matthree {1}{0}{0}{0}{-1}{0}{0}{0}{0}  \\[2ex]
  \lambda_4 \equiv s_x^{02}&= \matthree {0}{0}{1}{0}{0}{0}{1}{0}{0} &
  \lambda_5 \equiv s_y^{02}&= \matthree {0}{0}{-i}{0}{0}{0}{i}{0}{0}&
  \lambda_6 \equiv s_x^{12}&= \matthree {0}{0}{0}{0}{0}{1}{0}{1}{0} \\[2ex]
  \lambda_7 \equiv s_y^{12}&= \matthree {0}{0}{0}{0}{0}{-i}{0}{i}{0}&
  \lambda_8 &= \frac{1}{\sqrt{3}} \matthree {1}{0}{0}{0}{1}{0}{0}{0}{-2} 
\end{align*}
\end{widetext}

They are the generators of the Lie algebra of the special unitary group \(\mathrm{SU}(3)\) and can be thought of as the natural extension of Pauli matrices (generators of the Lie algebra of the SU(2) group). For each qutrit (with basis states $\ket{0}, \ket{1}, \ket{2}$), we calibrate a set of microwave pulses that resonantly drive the $\ket{0}\leftrightarrow\ket{1}$-transition and a separate set of pulses to address the $\ket{1}\leftrightarrow\ket{2}$-transition, providing universal control over the qubit subspaces $\{ \ket{0}, \ket{1} \}$ and $\{ \ket{1}, \ket{2} \}$. Our microwave control pulses directly perform rotations that correspond to exponentiating Gell-Mann matrices $s_x^{01},s_y^{01},s_z^{01},s_x^{12},s_y^{12}$. The Z-rotation ($s_z^{01}$) is implemented as a virtual Z gate in software by adjusting the phases of subsequent microwave pulses in that subspace~\cite{McKay_2017}. We extend this technique to the 12 subspace to also obtain the following rotation that is not one of the Gell-Mann matrices but that is very useful for single qutrit control since it is a virtual rotation with negligible error:
\begin{equation}
\begin{centering}
s_z^{12} = \matthree {0}{0}{0}{0}{1}{0}{0}{0}{-1}
\end{centering}
\end{equation}

 In principle one could also drive the $\ket{0}\leftrightarrow\ket{2}$-transition to directly implement rotations corresponding to $s_x^{02},s_y^{02}$. While it would be worthwhile to add these rotations to the available gate-set to compile circuits with lower depth, these two-photon transitions are more challenging to address as they require high-power. Luckily, all rotations generated by the remaining Gell-Mann matrices can be constructed from our available operations, for example
 \begin{equation}
\begin{centering}
\label{sx02}
e^{-i\frac{\theta}{2} s_{x/y}^{02}}=e^{-i\frac{\pi}{2} s_{x}^{12}}\cdot e^{-i\frac{\theta}{2} s_{x/y}^{01}}\cdot e^{i\frac{\pi}{2} s_{x}^{12}}
\end{centering} 
\end{equation}
where the right most operator is the first to act on the state, so time goes from right to left. Similarly, $\lambda_8$ can be constructed from $s_z^{01}$ and $s_z^{12}$ 

We write a rotation in one of these subspaces as $\theta^{k}_j = e^{-i \frac{\theta}{2}s_{j}^{k}}$, where $k = \{01, 12\}$ defines the subspace of the rotation, $j = \{ x , y, z\}$ the rotation axis and $\theta$ the rotation angle. As an example, the two available x-rotations and their corresponding rotation matrices are 

\begin{eqnarray}
\theta^{01}_x = \begin{psmallmatrix}\cos{\theta/2} &-i \sin{\theta/2}&0\\ -i \sin{\theta/2} & \cos{\theta/2}&0\\0&0&1 \end{psmallmatrix},  \nonumber \\
\theta^{12}_x = \begin{psmallmatrix}
1 & 0 & 0 \\
 0 & \cos{\theta/2}   & -i \sin{\theta/2} \\ 
 0 &-i \sin{\theta/2} & \cos{\theta/2}\\ 
 \end{psmallmatrix} {}.
\end{eqnarray}

This notation, combined with some useful qutrit and experiment specific operations, is also adopted in circuit diagrams displayed in the main text and in this document. 

Our gate-set consists of all Z-rotations along an arbitrary angle, combined with the Clifford operations operations in the 01 and 12 subspace :
\begin{eqnarray}
\theta^{k}_{j}  \textnormal{ with } \theta = \{ \pi,-\pi, \frac{\pi}{2},\frac{-\pi}{2} \}, \nonumber \\
\textnormal{where } k = \{01, 12\} \textnormal{ and } j = \{ x , y, z\}.
\end{eqnarray}

Three convenient gates to describe qudit-logic, which can be constructed from our universal gate-set, are the $X$ and $Z$ gates

\begin{equation}
X \ket{i} = \ket{i + 1 \text{ mod } d}
\end{equation}
\begin{equation}
Z \ket{i} = \omega^i \ket{i},
\end{equation}
where $\omega = \exp( i 2\pi / d)$. 

and the Hadamard gate
\begin{equation}
H=\frac{1}{\sqrt{d}} \sum_{i, j} \omega^{i j}|i\rangle\langle j|
\end{equation}

\subsection{Two-Qutrit operations: Controlled-SUM Gate}
In general $d$-dimensional qudits, two-qudit controlled-SUM and controlled-phase gates can be defined using the Pauli $X$ and $Z$ gates
We have:
\begin{eqnarray}
U_{\textrm{CSUM}}^{01}=\sum_{n=1}^{d}|n\rangle\langle n| \otimes X^{n}\nonumber \\ 
 U_{\textrm{C}\phi}=\sum_{n=1}^{d}|n\rangle\langle n| \otimes Z^{n}
\end{eqnarray}
Here the superscript $01$ indicates that the controlled-SUM is applied with $Q_0$ as the control qudit and $Q_1$ as the target; such a label is not necessary for $U_{\textrm{C}\phi}$, which is symmetric between the two qudits.  The two gates are equivalent up to a single-qudit Hadamard gate $H$ on the second qudit:
\begin{equation}
    (I\otimes H^\dagger) U_{\textrm{C}\phi} (I\otimes H) =  U_{\textrm{CSUM}}^{01},
\end{equation}
where the qudit Hadamard gate is defined to transform the $Z$ gate into the $X$ gate under conjugation.
Reversing the order of $H$ and $H^\dagger$ yields a controlled-MINUS gate, and changing which qubit receives the conjugation interchanges the control and target.  The entangling gates $U_{\textrm{C}\phi}$, $U_{\textrm{CSUM}}^{01}$, $U_{\textrm{CSUM}}^{10}$, $U_{\textrm{CMIN}}^{01}$, and $U_{\textrm{CMIN}}^{10}$ are therefore all  equivalent up to local (single-qudit) operations.  

In our system, we directly implement the two-qutrit $U_{\textrm{C}\phi}$ gate by interspersing periods of evolution under the cross-Kerr Hamiltonian (Eq.~\ref{eq:SuppCrossKerr}) with single-qutrit gates.  Intuitively, evolution under the cross-Kerr Hamiltonian imparts phases to the two-qutrit states $\ket{11}$, $\ket{12}$, $\ket{21}$, and $\ket{22}$, with values determined by the coefficients $\alpha_{ij}$.  By interspersing this phase accumulation with single-qutrit pulses exchanging the various states, we can ensure that each state accumulates exactly the phase required for the controlled-phase gate.  

We present two methods for implementing the controlled-phase gate in the manner described above.  The first uses fewer single-qutrit pulses and is conceptually simpler, but is not dynamically decoupled from the cross-Kerr interaction with neighboring qutrits.  The second is dynamically decoupled and is the one used in the teleportation experiment.  

\begin{figure}[b]
\includegraphics[scale=1]{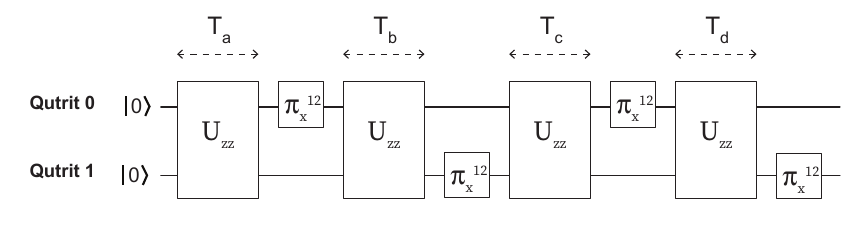}\caption{ \textbf{Four-segment pulse sequence implementing $U_{\textrm{C}\phi}$}. Four local $\pi$-pulses that effectively permute the eigenstates are interspersed with periods of free evolution $T_A, T_B, T_C, T_D$. The times depend on the cross-Kerr interaction parameters and are chosen such that the pulse sequence implements the desired diagonal unitary operation.
\label{fig:initcsum}}
\end{figure}

\textbf{First method:} 
As depicted in Fig.~\ref{fig:initcsum}, here we use four periods of cross-Kerr evolution, separated by pulses swapping the $\ket{1}$ and $\ket{2}$ states of a single qutrit.

  Denoting this swap pulse as $\pi^{12}_q$, where $q$ is the qutrit number, and evolution under the cross-Kerr Hamiltonian for a time $T$ as $ZZ_T$, the total pulse sequence is 
\begin{equation}
    ZZ_{T_A}~\cdot~\pi^{12}_0~\cdot~ZZ_{T_B}~\cdot~\pi^{12}_1~\cdot~ZZ_{T_C}~\cdot~\pi^{12}_0~\cdot~ZZ_{T_D}~\cdot~\pi^{12}_1.
\end{equation}

where the times $T_A$, $T_B$, $T_C$, $T_D$ depend on the cross-Kerr interaction parameters $\alpha_{ij}$.
For any choice of times, this operation imparts zero phase to the states $\ket{00}$, $\ket{01}$, $\ket{02}$, $\ket{20}$, $\ket{10}$, and non-zero relative phases $\phi_{11}$, $\phi_{12}$, $\phi_{21}$, and $\phi_{22}$ to the other basis states.  These phases are linear combinations of the delay times $T_A$, $T_B$, $T_C$, and $T_D$.  The transformation from delay times to induced phases is full rank (except for pathological values of the cross-Kerr coefficients), meaning that, given enough total delay time, this method can in principle generate an arbitrary two-qudit phase gate (the states that receive zero phase above can be made to gain arbitrary phase using only single-qutrit phase gates). On our particular chip, the coefficients $\alpha_{ij}$ allow us to implement the controlled-phase in this manner in roughly 1.5 $\mu$s for qutrit pairs $Q_1/Q_2$ and $Q_3/Q_4$.  

The drawbacks of this method become apparent when one tries to use it in a multi-qutrit algorithm.  If the two qutrits undergoing the controlled-phase are coupled to other qutrits via the same cross-Kerr Hamiltonian (as is the case for our chip), the above method will not work when the other qutrits are in superpositions of basis states, in which case entanglement  between them and the desired qutrits will be generated. The second method addresses this problem.

\begin{figure*}[t]
\includegraphics[scale=1]{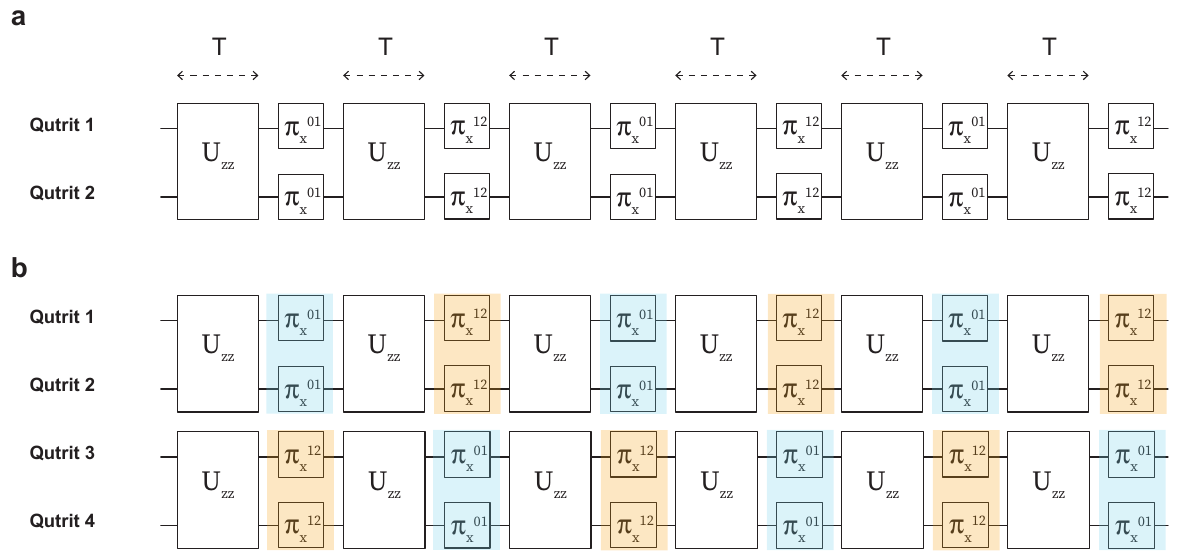}\caption{ \textbf{Six-segment pulse sequence implementing $U_{\textrm{C}\phi}$, dynamically-decoupled from static neighbors. } \textbf{(a)} By alternating six periods of free evolution with local permutation pulses, a diagonal phase gate can be implemented while also protecting the two qutrits from dephasing and static interactions with neighbors provided that they are static. \textbf{(b)} To perform two of these gates in parallel and maintain the decoupling, in this case between qutrit 2 and qutrit 3, the order of local permutations is swapped between pairs: qutrit 1 and qutrit 2 first are permuted in the  $\ket{0}$/$\ket{1}$ subspace (blue) while the first local operation on qutrit 3 and 4 is in the $\ket{1}$/$\ket{2}$ subspace (orange). This effectively decouples qutrit 2 and 3 in addition to performing the desired gates and decoupling qutrit 1 and 4 from their other neighbors (0 and 5).
\label{fig:csumv2}}
\end{figure*}

\textbf{Second method:} 
As depicted in Fig.~\ref{fig:csumv2} a, this method uses six \emph{equal} time periods of cross-Kerr evolution.  These are interspersed with single-qutrit pulses swapping the $\ket{0}$/$\ket{1}$ and $\ket{1}$/$\ket{2}$ subspaces, denoted $\pi^{12}_q$ and $\pi^{01}_q$, respectively.  The total pulse sequence consists of three repetitions of:
\begin{equation}
   \left[ ZZ_{T} ~\cdot~ \left(\pi^{12}_0 \otimes \pi^{12}_1\right) 
    ~\cdot~ ZZ_{T} ~\cdot~ \left(\pi^{01}_0 \otimes \pi^{01}_1\right) \right]
\end{equation}
For specificity, we have parameterized this pulse sequence with a single delay time, $T$; an appropriately chosen $T$ realizes the controlled-phase gate. The delay time $T$ is determined by the values of the cross-Kerr coefficients $\alpha_{ij}$ for each pair $Q_1/Q_2$ and $Q_3/Q_4$, and thus differs between the pairs; however, in practice, we find that a delay of 192 ns works well for both.

This pulse sequence constitutes a dynamically-decoupled implementation of the $U_{\textrm{C}\phi}$ unitary, as its operation is successful regardless of the states of the neighboring qutrits.  The dynamical decoupling arises because the single-qutrit pulses shuffle each qutrit's states $\ket{0}$, $\ket{1}$, and $\ket{2}$ such that an equal amount of time is spent in each state, regardless of the initial state of either qutrit. This shuffling `averages out' the cross-Kerr interaction with neighboring qutrits, such that no entanglement is generated.  

The particular teleportation algorithm we implement requires applying $U_{\textrm{C}\phi}$ on two pairs of qutrits $Q_1/Q_2$ and $Q_3/Q_4$. We use this dynamically decoupled pulse sequence for both pairs, and apply the gates simultaneously to reduce decoherence associated with a longer total gate time. Naively, the dynamical decoupling effect is weakened by this simultaneity, since the `neighboring qutrits', with respect to each individual pair, are no longer static.  Fortunately, we verify both theoretically and empirically that we can nevertheless decouple the unwanted interaction by reversing the order of the $\Pi^{12}$ and $\Pi^{01}$ gates between the two pairs (Fig.~\ref{fig:csumv2} b).  

\section{Dynamically-decoupled EPR preparation}
\label{app:DD}
\begin{figure*}
\includegraphics[width=0.9\textwidth]{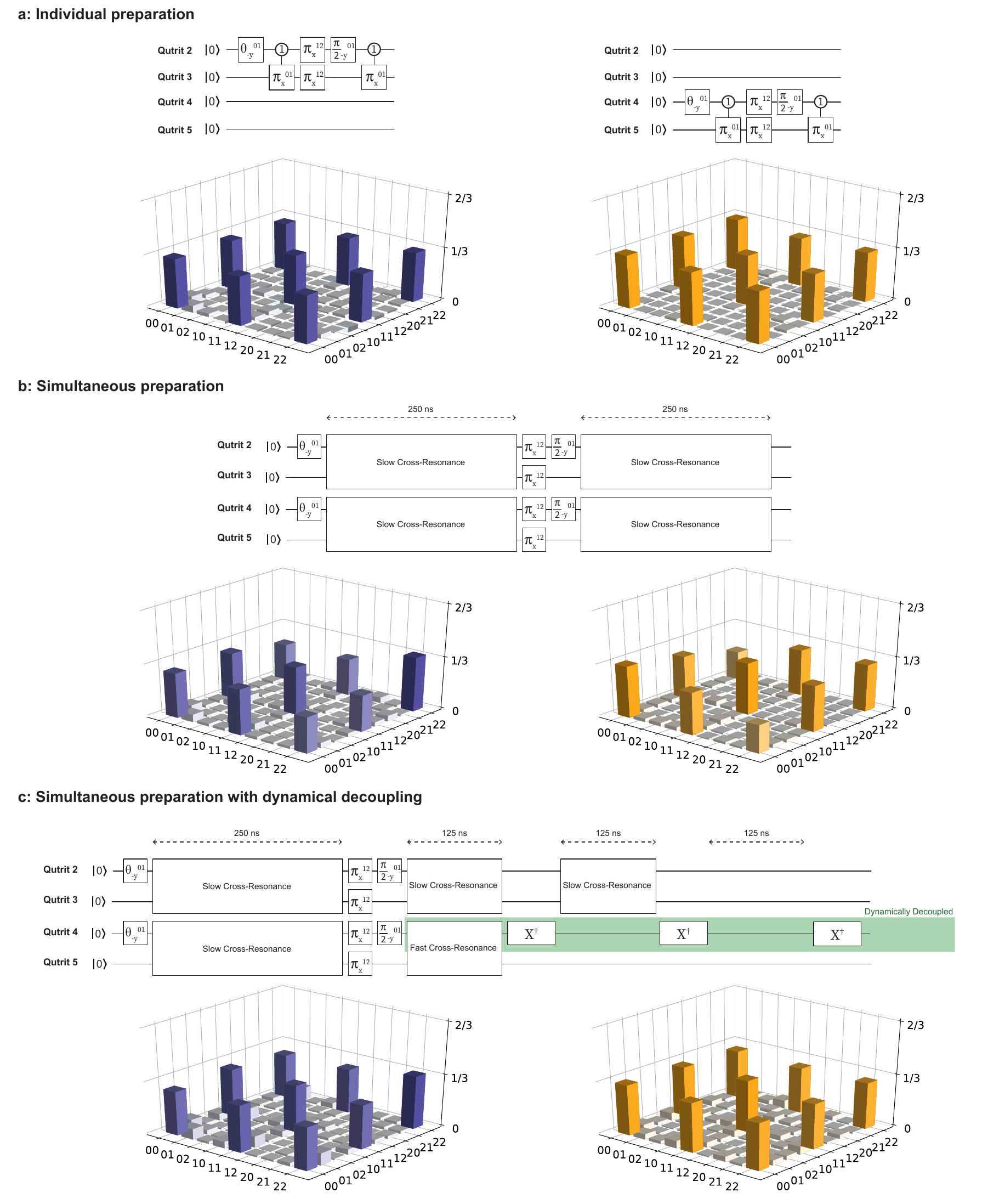}
\caption{ \textbf{Dynamically-decoupling the EPR pair preparation |} \textbf{a,} Density matrices of the $Q_2/Q_3$ (left, purple) and $Q_4$/$Q_5$ (right, orange) EPR pairs, prepared individually (i.e. with all other qutrits in the ground state). State fidelities for this dataset are $0.94 \pm 0.002$ and  $0.98 \pm 0.002$  respectively.  \textbf{b,} Density matrices of the same EPR pairs when prepared simultaneously without any dynamical decoupling.  Fidelites are markedly lower in this case,  $0.81 \pm 0.002$  and  $0.82 \pm 0.002$  respectively for the $Q_2/Q_3$ and $Q_4$/$Q_5$ pairs.  As discussed in the text, the loss of fidelity is due to unwanted entanglement arising from the cross-Kerr interaction between the two EPR pairs. \textbf{c,} EPR pairs prepared simultaneously using dynamical decoupling, with fidelities  $0.88 \pm 0.002$  and  $0.92 \pm 0.002$ , respectively.  
\label{fig:ddEPR}}
\end{figure*}

We prepare the two intial EPR pairs of the teleportation algorithm using the controlled-$\pi$ gate as discussed in the main text.  The basic sequence is presented in Fig. \ref{fig:csumv2}, and serves to prepare an EPR pair on either $Q_2$/$Q_3$ or $Q_4$/$Q_5$ individually, while all other qutrits are in the ground state $\ket{0}$.  Simultaneous EPR pair preparation, as required by the algorithm, necessitates a more complicated sequence that incorporates dynamical decoupling.  This necessity is demonstrated by Fig.~\ref{fig:ddEPR}(a-b), which compares the result of individual EPR preparation to joint EPR preparation without dynamical decoupling.  Joint preparation fidelities are much lower than those of individual preparation. From the measured density matrices, this loss seems to be largely due to a decrease in the off-diagonal elements (i.e. the coherences).  

To understand the source of this decrease in coherence, we measured the density matrix of the $Q_2$/$Q_3$ EPR pair while projecting the neighboring qutrit, $Q_4$, into each of its basis states.  Each of the three conditional density matrices we obtained was much purer (i.e. had much higher coherence) than the unconditional density matrix; however, the phases of each coherence differed depending on the state of $Q_4$.  These measurements suggest that the source of the decoherence was indeed unwanted entanglement between $Q_3$ and $Q_4$ arising from the cross-Kerr interaction.  

Qutrit State Tomography \cite{bianchetti_PRL_2010} after each step of the EPR preparation sequence allows us to pinpoint the portions of the sequence that contribute most strongly to the unwanted entanglement.  The cross-Kerr interaction affects the $\ket{2}$ states most strongly, and we find correspondingly that most of the entanglement occurs after the $\ket{2}$ state of $Q_3$ gets populated.  We take advantage of this by only dynamical decoupling the cross-Kerr interaction  after this point. As shown in Fig~\ref{fig:ddEPR}, the initial preparation of Bell states $(\ket{00} + \ket{11})/\sqrt{2}$, which does not involve the state $\ket{2}$, is performed without dynamical decoupling to reduce the error associated with additional single-qutrit gates.   

The mechanism underlying our decoupling sequence is most easily understood by first considering a simpler problem, of decoupling an unwanted cross-Kerr interaction between two qutrits during an idling period.  This can be accomplished by splitting the idling time into three equal time periods, and applying single-qutrit $X$ gates to one of the qutrits between each of the periods.  This shuffling of the populations decouples the entangling interaction into a product of local $Z$ interactions.  Using the same principle, we divide the controlled-$\pi$ operations in the relevant portion of simultaneous EPR preparation into three equal periods of 125 ns, and apply qutrit $X$ gates on $Q_4$ in between the waiting periods. These local gates on $Q_4$ do not commute with the controlled-$\pi$ operation on $Q_4$/$Q_5$. We therefore tune up a faster cross-resonance pulse that realizes the full controlled-$\pi$ gate between $Q_4$ and $Q_5$ within the first of the three 125 ns periods. The decoupling pulses on $Q_4$ can then be applied without interfering with the preceding entangling operation.  This sequence enables simultaneous EPR pair preparation with fidelities $0.88 \pm 0.002$  and  $0.92 \pm 0.002$ on $Q_2$/$Q_3$ and $Q_4$/$Q_5$, respectively.  

\bibliography{references}
\end{document}